\documentclass[a4paper,12 pt]{report}
\usepackage{rprt_m2,ref_thesis}
\usepackage{graphicx,latexsym}
\usepackage{amsmath,subfig}

\usepackage{amssymb}
\usepackage{amsthm}
\usepackage{amsfonts}
\usepackage{amscd}
\usepackage{graphics}
\usepackage{lscape}
\usepackage{rotating}

\DeclareMathSizes{80}{10}{12}{8}

\begin{document}
\theoremstyle{definition}
\pagestyle{repheadings} \baselineskip 0.3in
\parskip 0.0in
\begin{preamble}
\title{\Large \bf On the Fourier Transform Approach to Quantum Error Control}
\me
\author{Hari Dilip Kumar}

\maketitle
\parskip=.1in
\baselineskip=16pt \setcounter{page}{-2}
\newtheorem{thm}{Theorem}
\newtheorem{dfn}{Definition}
\newtheorem{fact}{Fact}

\begin{acknowledgement}
{Acknowledgements}

Firstly, I thank my parents for seeing me through the process of getting this degree. Next, I thank my guide, Prof. B. Sundar Rajan, for his patience and guidance. I thank my other teachers in IISc, from whom I learned so much. Finally, I thank all my friends, for many conversations and much laughter.
  
\end{acknowledgement}
\baselineskip 1.4pc \setcounter{page}{1}

\begin{abstract}{Preface}

	Quantum mechanics is the physics of the very small. Quantum computers are devices that utilize the power of quantum mechanics for their computational primitives. Associated to each quantum system is an abstract space known as the Hilbert space. A subspace of the Hilbert space is known as a quantum code. Quantum codes allow to protect the computational state of a quantum computer against decoherence errors.

	The well-known classes of quantum codes are stabilizer or additive codes, non-additive codes and Clifford codes. This thesis aims at demonstrating a general approach to the construction of the various classes of quantum codes. The framework utilized is the Fourier transform over finite groups.

	The thesis is divided into four chapters. The first chapter is an introduction to basic quantum mechanics, quantum computation and quantum noise. It lays the foundation for an understanding of quantum error correction theory in the next chapter.

	The second chapter introduces the basic theory behind quantum error correction. Also, the various classes and constructions of active quantum error-control codes are introduced.

	The third chapter introduces the Fourier transform over finite groups, and shows how it may be used to construct all the known classes of quantum codes, as well as a class of quantum codes as yet unpublished in the literature. The transform domain approach was originally introduced in \cite{krp}. In that paper, not all the classes of quantum codes were introduced. We elaborate on this work to introduce the other classes of quantum codes, along with a new class of codes, codes from idempotents in the transform domain.
    
	The fourth chapter details the computer programs that were used to generate and test for the various code classes. Code was written in the GAP (Groups, Algorithms, Programming) computer algebra package.
   
	The fifth and final chapter concludes, with possible directions for future work.

References cited in the thesis are attached at the end of the thesis.
\end{abstract}

\setcounter{page}{3} \tableofcontents 
\end{preamble}
\parskip=.1in
\baselineskip=16pt

\newcommand{\ud}{\mathrm{d}}
\newcommand{\drop}[1]{}

\begin{notation}
{Notation}
\begin{tabbing}
\hspace{50pt}\=\kill
$\mathbb{C}$ \> The complex numbers \\
$X$ \> The Pauli $X$ operator $\left( \begin{array}{cc}
0 & 1 \\ 
1 & 0
\end{array} 
\right) $ \\

$Y$ \> The Pauli $Y$ operator $\left( \begin{array}{cc}
0 & -i \\ 
i & 0
\end{array} 
\right) $ \\

$Z$ \> The Pauli $Z$ operator $\left( \begin{array}{cc}
1 & 0 \\ 
0 & -1
\end{array} 
\right) $ \\
$G_{n}$ \> The $n$-qubit Pauli group \\
$H$ \> The Hilbert space of the quantum system \\
$B(H)$ \> The space of linear operators on $H$ \\

$|\psi \rangle$ \> A ket in the Hilbert space \\
$\rho$ \> A density matrix \\
$U(d)$ \> The group of unitary matrices of size $d \times d$ \\
$\mathbb{C} S$ \> The group algebra of sums $\left\lbrace \sum_{s \in S} T_{s} s |T_{s} \in \mathbb{C} \right\rbrace $ \\
$\left\lbrace \rho_{i} \right\rbrace $ \> The set of irreducible representations of a group $S$ over $\mathbb{C}$ \\
Irr($S$) \> The set of irreducible characters of group $S$ over $\mathbb{C}$ \\
GL($W$) \> The set of automorphisms of a complex vector space $W$ \\
$P_{X+}$ \> The 1-dimensional projector onto the +1 eigenspace of the Pauli $X$ operator, etc\\

$E$ \> Error group
\end{tabbing} 

\end{notation}
\listoffigures

\chapter{Introduction}
\section{Introduction}

Quantum mechanics is the physics of the very small. Quantum computation and quantum information is the study of information processing tasks that can be accomplished using quantum mechanical systems \cite{nielsen}. A quantum computer is a device that uses quantum mechanical phenomena (such as entanglement and superposition) to effect operations on encoded data. Classical computers are the traditional computing devices that do not use these effects for computation.

The development of quantum computations follows an idea developed by Richard Feynman in 1982, that there seem to be essential difficulties in simulating quantum mechanical systems on classical computers, and that building computers based on the principles of quantum mechanics would allow us to avoid these difficulties. Hence one application of a quantum computer is the efficient simulation of quantum mechanical systems. Another strong incentive to build quantum computers is the existence of ``quantum algorithms" that run much faster than their classical counterparts. One such algorithm is the Shor algorithm for factoring numbers, which takes $O(b^{3})$ time to factor a $b$-bit number. The best classical algorithm, the General Number Field Sieve (GNFS) takes, by contrast $O(exp(\frac{64}{9}b)^{1/3} log (b)^{2/3})$ time to factor a $b$-bit number.

A formidable obstacle to building a quantum computer is the presence of decoherence, or quantum noise, that corrupts the state of the quantum memory. A theory of quantum error control has been built up to attempt to combat decoherence\cite{shor}. The method of quantum error control may be active error correction or passive error avoidance. In this thesis, we deal only with active error correction codes. 

The contributions of this thesis are
\begin{itemize}
\item A general mathematical framework, based on the Fourier transform over finite groups is presented. All the well-known classes of quantum error correction codes may be described in this framework. The Fourier transform approach was originally introduced in \cite{krp} for the case of error groups with Abelian index groups. However, the transform considered in this thesis is more general and includes error groups with non-Abelian index groups, and the codes obtainable therefrom.
\item The transform domain approach to designing quantum codes is highlighted. This allows for the potential discovery of new classes of codes
\item The error detection conditions for direct sum of Clifford codes have been investigated.
\item Some computer investigations have been carried out for a new class of codes - ``codes from idempotents in the transform domain".
\end{itemize}

In this chapter, the basics of quantum mechanics and quantum noise are presented. An introduction to classical coding theory is also included, as quantum error correction may in some cases be viewed as a classical coding problem.

\section{Quantum Mechanics}
This section contains a basic introduction to the quantum mechanics of finite dimensional systems, such as those used in quantum computation. The treatment is adapted from \cite{nielsen}. 

\subsection{Bra-Ket Notation}
In this section, the bra-ket notation for vectors in the Hilbert space is introduced. The standard quantum mechanical notation for a vector in a vector space is
\[
	|\psi \rangle.
\]
$\psi$ is a label for the vector; any label is permissible. Every ket $|\psi \rangle $ has a dual bra, written as $\langle \psi |$. The dual bra is a linear functional from the Hilbert space to the complex numbers, $\mathbb{C}$. With bras and kets, we may form inner products such as
\[
\langle \psi | \phi \rangle
\]
and also rank one operators such as
\[
|\psi \rangle \langle \phi |.
\]
The above operator maps the ket $|\rho \rangle$ to the ket $| \psi \rangle \langle \phi | \rho \rangle$.

\subsection{The Postulates of Quantum Mechanics}
The postulates of quantum mechanics are now presented. The postulates provide a connection between the physical world and the mathematical formalism of quantum mechanics.

\subsubsection*{Postulate 1: State Space}
Associated to any isolated physical system is a complex vector space with inner product (a Hilbert space) known as the state space of the system. The system is completely defined by its state vector, which is a unit vector in the system's state space.
A 2-dimensional state space is popularly referred to as a ``qubit" (quantum bit) in the literature. The state spaces used in quantum computation are usually finite-dimensional.

\subsubsection*{Postulate 2: Time Evolution}
The evolution of a closed quantum system is described by a unitary transformation. That is, the state $|\psi \rangle$ of a system at time $t_{1}$ is related to the state $|\psi '\rangle$ of the system at time $t_{2}$ by a unitary operator $U$ that depends only on the times $t_{1}$ and $t_{2}$. The evolution is given by 
\[
\mid \psi '\rangle = U \mid \psi \rangle.
\]

\subsubsection*{Postulate 3: Projective Measurements}
Although more general measurements are possible in quantum mechanics, they are not widely used in quantum error correction. In this thesis only projective measurements are used. 
A projective measurement is described by an observable, $M$, a Hermitian operator on the state-space of the system being observed. The observable has a spectral decomposition
\[
M = \sum_{m} mP_{m},
\]

\noindent where $P_{m}$ is the projector onto the eigenspace of $M$ with eigenvalue $m$. The possible outcomes of the measurement correspond to the eigenvalues, $m$, of the observable. Upon measuring the state $| \psi \rangle$, the probability of getting the result $m$ is
\[
p(m) = \langle \psi |P_{m}| \psi \rangle.
\]

Given that the outcome $m$ occurred, the state of the quantum system immediately after measurement is
\[
\frac{P_{m} | \psi \rangle}{\sqrt{p(m)}}.
\]

\subsubsection*{Postulate 4: Composite Quantum Systems}
The state space of a composite quantum system is the tensor product of the state spaces of the component physical systems. If there are $n$ systems, with states $|\psi_{i} \rangle$, the composite system has the state $|\psi_{1} \rangle \otimes | \psi_{2} \rangle \otimes ... \otimes | \psi_{n} \rangle $.

\subsection{The Density Operator and Ensembles of Quantum States}
The density operator is also called the density matrix. It is a tool used to track the preparation history of a mixture of states in quantum mechanics. Suppose a quantum system is in one of a number of states $|\psi_{i}\rangle$ where $i$ is an index, with respective probabilities $p_{i}$. The density operator for the system is defined by the equation
\[
\rho = \sum p_{i} | \psi_{i}\rangle  \langle \psi_i|.
\]
The state $I/2$, for example, represents a completely mixed ensemble.
The density operator is important because it offers a natural tool for describing the effect of environmental noise on open quantum systems. The map from ensembles to density operators is not injective; this is captured by the notion of ``unitary freedom" in the density operator. Another characterization of density operators is the following: An operator $\rho$ is a density operator if, and only if, it satisfies the trace and positivity conditions: the trace of $\rho$ should be 1, and $\rho$ should be a positive operator.

\section{Quantum Noise and Quantum Operations}
In this section the main formalism for dealing with quantum noise is introduced. The notation $H$ is used for the system Hilbert space, and $B(H)$ represents the space of bounded linear operators on this Hilbert space.

\subsection{Operator-Sum Representation of Quantum Operations}
Suppose a quantum system is in a state $\rho$, and undergoes time-evolution, possibly in the presence of an external environment, to the state $\rho '$. The state change is given by
\[
\rho ' = E(\rho ).
\]

In this case the map $E :B(H) \rightarrow B(H)$ is called a quantum operation or channel.
For a restricted class of maps useful in quantum error correction (completely positive maps), the quantum operation $E(\rho )$ possesses an operator-sum representation, considerably simplifying the mathematical formalism. The operator-sum notation is
\[
\rho' =E(\rho) = \sum_{k} E_{k} \rho E_{k}^{\dagger},
\]

\noindent where each $E_{k} \in B(H)$. The operators  $\{ E_{k} \}$ are known as operation elements for the quantum operation $E$. We also require that
\[
\sum E_{k}^{\dagger} E_{k} = I,
\]
an additional property that reflects the trace-preserving property of the map $E$.
\subsubsection*{Physical Interpretation of the Operator-Sum Notation}
In the operator-sum notation $\rho' = \sum E_{k} \rho E_{k}'$ the action of the quantum operation is equivalent to taking the state $\rho$ and randomly replacing it by ${E_{k}\rho E_{k}^{\dagger}}/{Trace(E_{k} \rho E_{k}^\dagger)}$ with probability $Trace(E_{k} \rho E_{k}^{\dagger})$. Hence the complete positivity of the map $E$ reflects the fact that density matrices are taken to density matrices, and the trace-preservation of $E$ represents the conservation of total probability.
\subsection{Depolarizing Channel}
The depolarizing channel is the type of noise considered in this thesis. The depolarizing channel acts on quantum states $\rho$ by the transformation

\[
E(\rho) = \frac{pI}{2} + (1-p)\rho.
\]

\noindent That is, with probability $p$, the state is completely mixed, and with probability $1-p$, the state is untouched. The above identity can also be written, using
\[
I/2= \frac{\rho+X\rho X+ Y\rho Y+ Z\rho Z}{4},
\]
\noindent where $X=\left( \begin{array}{cc}
0 & 1 \\ 
1 & 0
\end{array}\right) $,
$Y=\left( \begin{array}{cc}
0 & -1 \\ 
1 & 0
\end{array}\right) $,
$Z=\left( \begin{array}{cc}
1 & 0 \\ 
0 & -1
\end{array}\right) $,
are the Pauli operators,
as
\[
E(\rho) = (1-p) \rho + (p/3)(X\rho X +Y\rho Y + Z\rho Z.
\]

\noindent This has the interpretation that the state $\rho$ is left alone with probability $1-p$ and the operations $X, Y$ and $Z$ are each applied with probability $p/3$.

\section{Introduction to (Classical) Coding Theory}
This section introduces classical coding theory for binary systems in brief. An insight into classical coding theory is useful for quantum coding, especially in the case of ``CSS" and ``additive" quantum codes.

The essential components of a communication system are illustrated above.
\begin{figure}[!htb]
\includegraphics[scale =1.2]{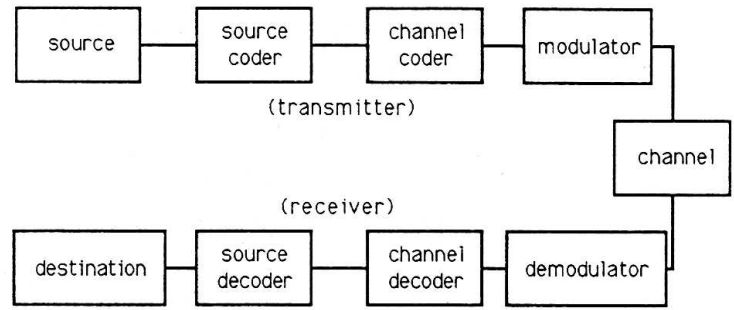} 
\caption {Elements of a Communication System}
\label{comm} 
\end{figure}

The simplest channel model usually asumed for coding is the Binary Symmetric Channel (BSC), illustrated below

\begin{figure}[!htb]
\includegraphics[scale =0.9]{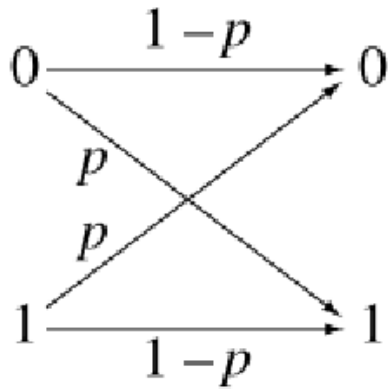} 
\caption {Binary Symmetric Channel}
\label{bsc} 
\end{figure}

With probability $1-p$, the state $s \in \left\lbrace 0,1 \right\rbrace $ remains unchanged; with probability $p$ the state changes.

This probabalistic channel model causes data to be corrupted as it is sent across the channel. This can be mitigated to some extent by using a binary error correcting code. The simplest example of an error correcting code is the repetition code, given by
\[
0 \mapsto 000
\]
\[
1 \mapsto 111
\]
A majority decision is taken at the receiver side to determine the actual bit sent. This simple code can protect for one bit of error in the transmission. The code is denoted as a $[3,1,3]$ error correcting code. More generally, 

\begin{dfn}
an $[n,k,d]$ binary linear code may be defined as a linear subspace, $C$, of $F_{2}^{n}$, of rank $k$, with the ability to correct $\lfloor \frac{d-1}{2} \rfloor$ errors, and detect $d-1$ errors.
\end{dfn}

A binary linear code may also be defined by its parity check matrix over the finite field $GF(2)$.
\begin{dfn}
Let $H$ be any binary matrix. The linear code with parity check matrix $H$ consists of all vectors $x$ such that
\[
Hx^{tr} = 0.
\]
\end{dfn}

Associated with this subspace of dimension $k$ is a matrix $G$, the generator matrix for the code. The generator matrix is given by
\[
G = \left( \begin{matrix}
v_{1} \\ 
v_{2} \\ 
 \\ 
v_{k}
\end{matrix} 
\right) 
,
\]
where $v_{1}, ...,v_{k}$ are basis vectors for the code space $C$. The generator matrix is an $k \times n$ sized matrix over $F_{2}$
and the parity check matrix is an $n \times (n-k)$ matrix over the finite field.

\section{Scope of This Thesis}
Quantum error control deals with how to mitigate the effects of noise upon a quantum system (say a quantum memory for a computer). The error control may be active error correction, passive error avoidance, or a combination of the two. 

In quantum error correction, the state of the computer is actively restricted to a subspace of the system Hilbert space. Passive error avoidance is examplified by methods such as noiseless subsystems (NS) \cite{ns}, and decoherence free subspaces (DFS) \cite{dfs}. A combination of the active and passive approaches is found in Operator Quantum Error Correction (OQEC) \cite{oqec}. In this thesis, we focus exclusively on active error control or subspace encoding error correction schemes. For such codes, a very general framework for code construction is described. It is an open problem to generalize the results beyond active error control schemes, to Operator Quantum Error Correction.

\chapter{Quantum Coding: Theory and Constructions}

\section{Introduction}
The previous chapter introduced the basics of quantum computation, quantum noise and classical coding theory. In this chapter, this foundation is built upon to develop the theory of quantum error correction. Once this general theory has been introduced, the various classes of active quantum error correcting codes are presented. This sets the stage for the next chapter, where all these classes are analyzed in light of the Fourier transform over finite groups.

\section{The Case Against Quantum Error Correction}
Noise is a great bane of information processing systems. To protect classical information-processing systems from classical noise, the theory of error correcting codes \cite{sloane} was developed. This was introduced in the first chapter. Developing a similar theory for protecting quantum states from quantum noise faces the following serious obstacles.
\subsection{Quantum No-Cloning Theorem}
According to quantum mechanics, it is impossible to build a machine that will duplicate an arbitrary quantum state \cite{wz}. This is called the quantum no-cloning theorem. Hence the equivalent of repetition codes seem to be disallowed in quantum mechanics, as there is no way to perform the mapping
\[
| \psi \rangle \longmapsto | \psi \rangle | \psi \rangle | \psi \rangle.
\]
\subsection{Errors Are Continuous}
For a quantum mechanical system, such as a 2-dimensional qubit, with state space $H=\mathbb{C} ^2$, the noise is in general any matrix from $B(H)$, the space of linear operators on the Hilbert space. This is a continuous space, hence there are an infinite number of possibilities for the noise.

\subsection{Measurement Destroys Quantum Information}
In classical error correction, the output of the channel is observed and decoding is performed based on this. Observation in quantum mechanics generally destroys the quantum state under observation and makes recovery impossible.

\section{Quantum Error Correcting Codes}

In spite of the above-mentioned serious obstacles, quantum error correction is still possible. This can be illustrated by the Shor code, the first quantum code to be found. In general, a quantum code is a subspace of the Hilbert space of the quantum system used to protect quantum information, Quantum codes are denoted by $((n, K, d))_{q}$ if they encode a $K$-dimensional subspace in a $q^{n}$ dimensional space, and correct errors of weight upto $\lfloor \frac{d-1}{2} \rfloor$ tensor factors. A specialized notation for stabilizer codes is $[[n,k,d]]_{q}$ where a $q^{k}$ dimensional code space is now used.
\subsection{The Shor Code}
The Shor code is a $[[9,1,3]]$ code that corrects for one arbitrary error on any qubit. The Shor code is given by the logical basis states
\[
|0_{L} \rangle = (|000 \rangle + |111 \rangle)(|000 \rangle + |111 \rangle)(|000 \rangle + |111 \rangle),
\]
\[
|1_{L} \rangle = (|000 \rangle - |111 \rangle)(|000 \rangle - |111 \rangle)(|000 \rangle - |111 \rangle).
\]

To detect a bit flip in the first group of 3 qubits, the measurements $Z_{1}Z_{2}$ and $Z_{2}Z_{3}$ are performed. These yield 2 classical bits of information that specify if one of the first three qubits has undergone a bit flip error. 
\begin{center}
\begin{tabular}{|c|c|c|}
\hline $Z_{1}Z_{2}$ & $Z_{2}Z_{3}$ & \\ 
\hline +1 & +1 & No error \\ 
\hline +1 & -1 & $X_{3}$ \\ 
\hline -1 & +1 & $X_{1}$ \\ 
\hline -1 & -1 & $X_{2}$ \\ 
\hline 
\end{tabular} 
\end{center}
Similar measurements can be performed on the other groups of 3 qubits. Similarly, phase flip errors affecting the clusters of 3 qubits may be detected by the operators $X_{1}X_{2}...X_{6}$ and $X_{4}X_{5}...X_{9}$.

Measuring all these operators allows to correct for the depolarizing channel at a single arbitrary qubit location. However, correction for this restricted model is sufficient for correction of arbitrary errors on the qubit. While this can be formally proved using the Knill-Laflamme conditions as a starting point \cite{nielsen}, an intuitive explanation is given here.
Suppose the state of the system after noise acts is
\[
E(\rho) = \sum E_{i} \rho E_{i}^{\dagger}.
\]

A state $|\psi \rangle$ of the system is taken to
$\sum E_{i} | \psi \rangle  \langle \psi | E_{i}^{\dagger}$.

It is assumed that some particular $E_{i}$ acts only on one qubit.
Focussing on this $E_{i}$, we write out

$E_{i} = \alpha_{1} I + \alpha_{2} X + \alpha_{3} Y + \alpha_{4} Z$,

\noindent where $X, Y$ and $Z$ are the Pauli operators. This is possible because the matrices $I, X, Y$ and $Z$ form a basis for the operator space $B(H)=M_{2}(\mathbb{C})$, where $H=\mathbb{C}^{2}$ is the Hilbert space of one qubit.
The state $E_{i}| \psi\rangle$ is therefore
\[
\alpha_{1}I |\psi\rangle + \alpha_{2} X |\psi \rangle + \alpha_{3} Y |\psi\rangle + \alpha_{4} Z |\psi \rangle
\]
\noindent which is a superposition of the four states $\alpha_{1} I |\psi \rangle $, $ \alpha_{2} X|\psi \rangle$, $ \alpha_{3}Y | \psi \rangle$,$\alpha_{4} Z |\psi \rangle$. Measurement of the error syndrome causes the state to collapse to one of the four states $|\psi \rangle, X|\psi \rangle, Z|\psi \rangle$ or $XZ| \psi \rangle$, from which recovery may be performed by the appropriate inversion.
This error correction (measure and rotate) results in the original state $|\psi \rangle$ being recovered, despite the fact that the initial error was arbitrary.
\section{Quantum Error Correction Theory}
\subsection{Knill-Laflamme or Quantum Error Correction Conditions}
These fundamental conditions \cite{knill1} are also called the quantum error-correcting conditions. They allow us to test how well a QECC corrects for a particular error model. 

\begin{thm}\cite{knill1,nielsen}
Let $C$ be a quantum code and let $P$ be the projector onto $C$. Suppose $E$ is a quantum operation with operation elements $\left\lbrace E_{i} \right\rbrace $. A neccessary and sufficient condition for the existence of an error correction operation $R$ correcting $E$ on $C$ is that
\[
PE_{i}^{\dagger}E_{j}P = \alpha_{ij}P \  \ \ \forall E_{i}, E_{j}
\]
for some Hermitian matrix $\alpha$ of complex numbers.
\end{thm}
This condition can equivalently be formulated as
\[
\langle \psi_{a}|E_{i}E_{j} | \psi_{b} \rangle = \delta_{ab}\alpha_{ij}
\]
for logical basis states $\left\lbrace \psi_{i} \right\rbrace $ of the code space.

The operation elements $\left\lbrace E_{i} \right\rbrace $ for the noise $E$ are called errors, and if such an $R$ exists, $\left\lbrace E_{i} \right\rbrace $ is called a correctable set of errors. The proof of this theorem is constructive; the recovery operation is explicitly constructed using projective (syndrome) measurements and unitary rotations.
Also, there are conditions for the detectability of errors: An error operator $D$ is detectable by a code with projector $P$, if, and only if,
\[
PDP = \lambda_{D} P
\]
for some $\lambda_{D} \in \mathbb{C}$ \cite{clifford}.
\subsection{Discretization of Errors}
The following theorem shows how it is enough to correct for a basis of the error space acting on the system. This allows to correct for a discrete and finite set of errors, even though the error space is infinite. 
\begin{thm}\cite{nielsen}
Suppose $C$ is a quantum code and $R$ is the recovery operation to recover from a noise process $E$ with elements $\left\lbrace E_{i} \right\rbrace $. Suppose $F$ is a quantum operation with operation elements $\left\lbrace F_{j} \right\rbrace $ which are linear combinations of the $E_{i}$. Then the error correction operation $R$ also corrects for the effects of the noise process $F$ on the code $C$.
\end{thm}

This result follows basically from the linearity of the Knill-Laflamme conditions in the errors $E_{i}$.

\section{Error Bases}
Motivated by the above theory, we search for suitable bases for the error space acting on the system.
\subsection {Pauli Basis}
The prototypical example of an error basis is the Pauli basis, consisting of the matrices $\left\lbrace I, X, Y, Z\right\rbrace$. These matrices form a basis for the vector space $B(H)=M_{2} (\mathbb{C})$. As shown in the previous section, if a quantum error correction code corrects for errors $\left\lbrace I, X, Y, Z\right\rbrace$ on a single qubit, then any arbitrary error on the single qubit can be corrected. The multiplicative closure of the Pauli matrices forms a finite group, the Pauli group $G_{1}$ on 1 qubit. An error basis for an $n$-qubit system may be formed by taking the $n$-fold tensor products of the Pauli matrices together, i.e. $\left\lbrace I, X, Y, Z \right\rbrace ^{\otimes n}$.
\subsection{Nice Error Bases}
 The properties of the Pauli basis are abstracted in the notion of ``nice error bases", The Pauli matrices, with identity, form a nice error basis. Nice error bases are defined as follows.

\begin{dfn}
Let $G$ be a finite group of square order $n^{2}$.
A nice error basis is a set of unitary $n \times n$ matrices $D(g)$ parametrized by the index group $G$ such that
\begin{itemize}
\item $D(1)$ is the identity matrix,
\item Trace $D(g) = 0$ for all nonidentity elements $g$ of $G$,
\item $D(g)D(h) = \omega(g,h) D(gh)$ where $\omega(g,h)$ is a phase factor. 
\end{itemize}

\end{dfn}

Nice unitary error bases are useful when coding over higher dimensional systems ("qudits"). They are also fundamental to the theory of Clifford coding for quantum error correction. The multiplicative closure of a nice error basis is called an error group, or abstract error group.

\section{Quantum Code Constructions}

\begin{figure}[h]   
\includegraphics[scale =0.5]{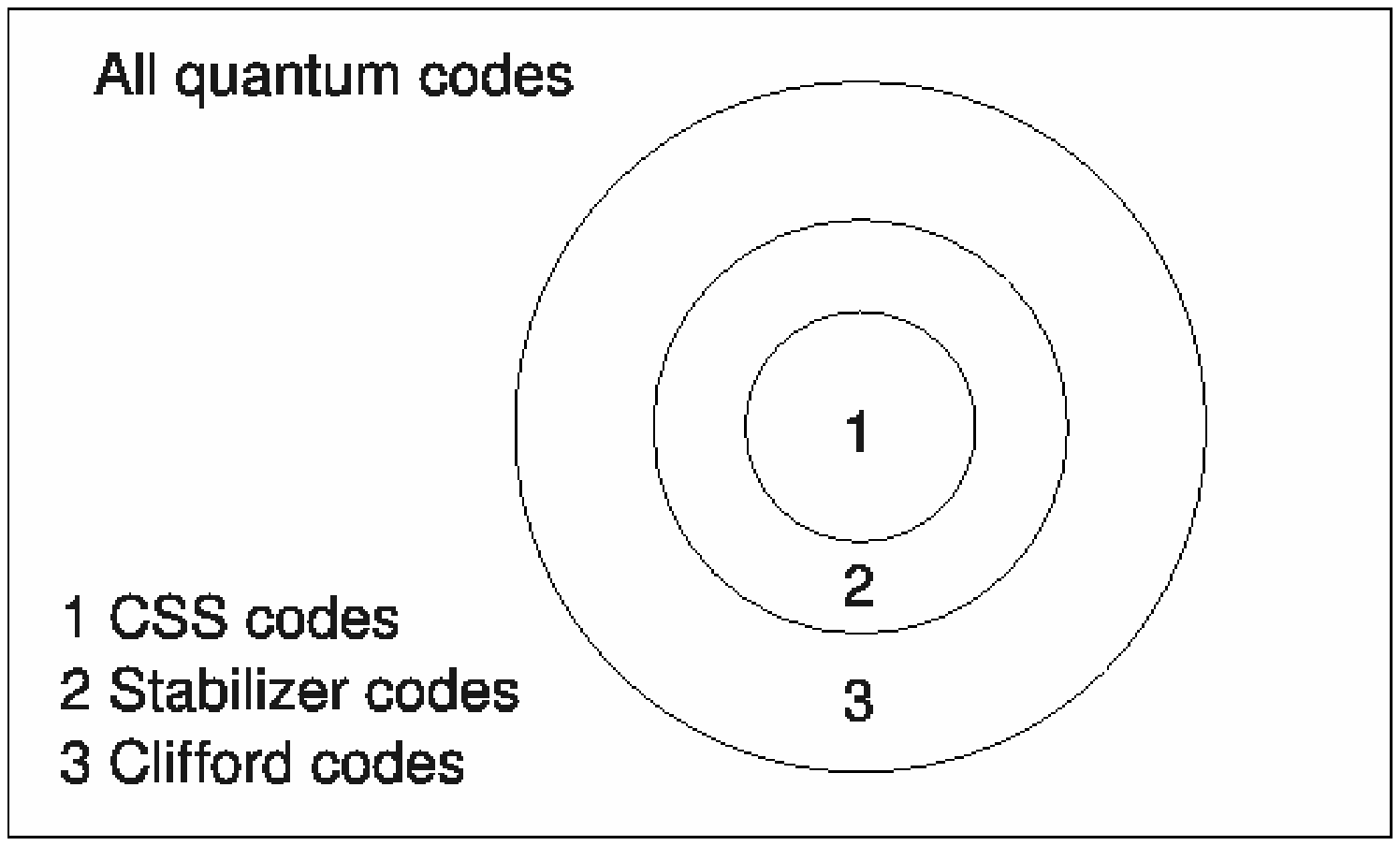} 
\caption {Some Classes of Quantum Codes}
\label{quant} 
\end{figure}
A quantum code in general is a subspace of the state space of the system. A quantum code is said to have minimum distance $d$ if it can correct all errors of weight upto $\lfloor \frac{d-1}{2} \rfloor$.

\subsection{CSS Codes}
CSS (Calderbank-Shor-Steane) codes are a large family of quantum codes that were introduced in \cite{css1,css2}. They are a subclass of the more general stabilizer codes.
\begin{fact}
Suppose $C_{1}$ and $C_{2}$ are $[n, k_{1}]$ and $[n, k_{2}]$ classical linear codes such that $C_{2} \leq C_{1}$ and $C_{1}$ and $C_{2}^{\perp}$ both correct $t$ errors. We can then define an $[[n, k_{1}-k_{2}]]$ quantum code CSS($C_{1},C_{2}$) capable of correcting errors on $t$ qubits.
\end{fact}

Say $x \in C_{1}$ is any codeword of the code $C_{1}$. Then the state $|x+C_{2} \rangle$ is defined by
\[
|x+C_{2} \rangle = \frac{1}{\sqrt{|C_{2}|S|}} \sum_{y \in C_{2}} |x+y \rangle
\]

\noindent where $+$ is addition modulo 2. The state $|x+C_{2}\rangle$ depends only upon the coset of $C_{2}$ that $x$ is in. CSS($C_{1}, C_{2}$) is an $[[n, k_{1}-k_{2}]]$ quantum code as there are $2^{k_{1}-k_{2}}$ cosets.
The classical error correcting properties of the codes $C_{1}$ and $C_{2}$ determine the error correcting properties of the quantum code CSS($C_{1},C_{2}$). The analysis is as follows.
Say $e_{1} \in F_{2}^{n}$ represents the bit-flip errors, and $e_{2} \in F_{2}^{n}$ represents the phase-flip errors. The quantum state $|x+C_{2}\rangle$ is now corrupted to
\[
\frac{1}{|C_{2}|} \sum_{y \in C_{2}} (-1)^{(x+y).e_{2}} |x+y+e_{1} \rangle.
\]

First, an ancilla system is used to measure the error syndrome for the bit flip errors. This produces the state
\[
\frac{1}{|C_{2}|} \sum_{y \in C_{2}} (-1)^{(x+y).e_{2}} |x+y+e_{1} \rangle |H_{1}e_{1} \rangle.
\]
Knowing the error syndrome, by measuring the ancilla, the state can be changed to
\[
\frac{1}{|C_{2}|} \sum_{y \in C_{2}} (-1)^{(x+y).e_{2}} |x+y \rangle 
\]
by applying NOT gates at the appropriate locations and discarding the ancilla.

To detect phase errors, the Hadamard gates are applied to each qubit, taking the state to
\[
\frac{1}{\sqrt{|C_{2}|2^{n}}}\sum_{z} \sum_{y \in C_{2}} (-1)^{(x+y).(e_{2}+z)} |z \rangle.
\]

\noindent where the sum is over all possible values for $n$-bit $z$. Setting $z' = z+e_{2}$, the state may be rewritten as

\[
\frac{1}{\sqrt{|C_{2}|2^{n}}}\sum_{z'} \sum_{y \in C_{2}} (-1)^{(x+y).(z')} |z '+e_{2} \rangle
\]

\[
= \frac{1}{\sqrt{|C_{2}|2^{n}}}\sum_{z'} \sum_{y \in C_{2}} (-1)^{(x+y).(z')} |z'+e_{2} \rangle
\]

\[
= \frac{1}{\sqrt{2^{n}/|C_{2}|}} \sum_{z' \in C_{2}^{\perp}} (-1)^{(x).(z')} |z' \rangle.
\]

This state appears identical to a bit flip error described by $e_{2}$. The same error correction process as before (use the parity check matrix $H_{2}$ and measure the ancilla) may be used to revert the state to
\[
\frac{1}{\sqrt{ 2^{n}/|C_{2}|}}\sum_{z' \in C_{2}^{\perp}} (-1)^{x.z'}|z' \rangle.
\]

The Hadamard gates are now applied again to each qubit, to return to the original state
\[
\frac{1}{\sqrt{|C_{2}|}} \sum_{y \in C_{2}} |x+y \rangle.
\]

CSS codes may be viewed as as a subclass of stabilizer codes. Measurement of the stabilizer generators to produce an error syndrome simplifies the recovery process for CSS codes.

A coding-theoretic formulation is obtained as follows, from \cite{salah}.

\begin{fact}{CSS Code Construction}
Let $C_{1}$ and $C_{2}$ denote two classical linear codes
with parameters $[n, k_{1}, d_{1}]_{q}$ and $[n, k_{2}, d_{2}]_{q}$ such that $C^{\perp}_{2} \leq C_{1}$. Then there exists
a $[[n, k_{1} + k_{2} - n, d]]_{q}$ stabilizer code with minimum distance $d = min \left\lbrace wt(c) | c \in (C_{1} \backslash C^{\perp}_{2} ) \cup (C_{2} \backslash C^{\perp}_{1} ) \right\rbrace  \geq min \left\lbrace d_{1}, d_{2} \right\rbrace .$
\end{fact}

\subsection{Stabilizer or Additive codes}
These codes are introduced in \cite{gottesma}. Stabilizer codes are a large class of codes that include CSS codes as a subclass. They are most easily explained using the qubit Pauli group $G_{n}$, although generalizations to higher dimensional Pauli groups \cite{knill} and other error bases \cite{beyond2} are straightforward.

The $n$-qubit Pauli group $G_{n}$ is formed as outlined in section 2.5. An Abelian subgroup, $S$, of this error group is chosen, not containing any nontrivial multiple of identity $\omega I$, $w \neq 1$. The subgroup $S$ needs to be commuting because only commuting operators can be simultaneously diagonalized. Non-trivial multiples of identity are disallowed because we wish the joint +1 eigenspace of the group to be nontrivial.
The stabilizer code $C$ is defined as the subspace of $H$ that is fixed by the group $S$, i.e. it is the joint +1 eigenspace of $S$.
\subsubsection*{Error Correction Properties of Stabilizer Codes}
The set of elements that commute with everything in $S$ is the centralizer $C(S)$. Because of the properties of the Pauli group $G_{n}$, $C(S) = N(S)$, the normalizer of $S$ in $G$. A quantum code with stabilizer $S$ will detect all errors that are either in $S$ or anticommute with some element in $S$ \cite{gott_thesis}. The code will correct any set of errors $\left\lbrace  E_{i} \right\rbrace $ if $E_{a}^{\dagger} E_{b} \in S \cup \left\lbrace G_{n} - N(S) \right\rbrace \forall E_{a}, E_{b}$. The code is said to be of distance $d$ if $N(S) \backslash S$ contains no elements of weight less than $d$. If $S$ has elements of weight less than $d$, the code is called degenerate.

If the generators of the code are $ \left\lbrace g_{1}, g_{2}, \ldots, g_{n-k} \right\rbrace $, the codespace has dimension $2^{n-k}$. This can be seen by taking the trace of the code projector: Trace $\frac{1}{|S|} \Pi (I + g_{i})$.

\subsection{Additive or GF(4) Formalism for Stabilizer codes}
Additive codes were introduced in \cite{crss}. Additive codes are synonymous with stabilizer codes, although there is a slight technical difference in their definition.
In this formalism, two finite spaces are used. The first is the binary vector space $GF(2)^{2n}$. Elements of $GF(2)^{2n}= \overline{E}$ are denoted by $(a|b), \ \ a,b \in GF(2)^{n}$. $\overline{E}$ is equipped with the inner product $((a|b),(a'|b'))=ab' + a'b$\ $\in GF(2)$. This is a symplectic inner product, meaning that $((a|b),(a|b))=0$. The following map is used from the Pauli group $G_{n}$ to the space $\overline{E}$
\[
\Psi: G_{n} \longrightarrow G_{n}/Z(G_{n}) = \overline{E}
\]
\[
i^{\lambda}X(a)Z(b) \longmapsto (a|b) \in \overline{E}
\]

\noindent to map elements of the Pauli group to the symplectic space.
The problem of finding an Abelian subgroup of $G_{n}$, containing the center $Z(G_{n})$ is equivalent under this map to finding a symplectically self-orthogonal space in $\overline{E}$. In short, the following theorem \cite{crss} holds true.
\begin{thm}\cite{crss}
Suppose $\overline{S}$ is an $n-k$ dimensional linear subspace of $\overline{E}$ which is contained in its dual $\overline{S^{\perp}}$ (with respect to the symplectic inner product), and is such that there are no vectors of weight $<d$ in $\overline{S^{\perp}} \backslash \overline{S}$. Then there is a quantum error correcting code mapping $k$ qubits to $n$ qubits which can correct $\lfloor \frac{d-1}{2} \rfloor$ errors.
\end{thm}
The Abelian group $S$ is obtained as the the lift $\Psi^{-1}(\overline{S})$. An eigenspace, for any chosen linear character $\chi$ of this group satisfying $\chi(iI) = i$, is chosen as the code space.
The problem of coding may also be expressed over $GF(4)$, the finite field with 4 elements ($\left\lbrace 0, 1, \omega, \overline{w} \right\rbrace $) via the map
\[
\phi: \overline{E} \longrightarrow GF(4)^{n}
\]
\[
(a|b) \longrightarrow \omega a + \overline{\omega}b
\]

\noindent where $\omega \in GF(4)$.

The symplectic inner product on $\overline{E}$ is now converted to the trace-Hermitian inner product, i.e. $((a|b),(a'|b')) = Trace(\phi(v), \overline{\phi(v')})$ where $v=(a|b)$ and $v'=(a'|b')$. The following theorem is then obtained.

\begin{thm}\cite{crss}
Suppose $C$ is an additive self-orthogonal subcode of $GF(4)^{n}$ containing $2^{n-k}$ vectors, such that there are no vectors of weight $<d$ in $C^{\perp} \ C$. Then any eigenspace of $\phi^{-1}(C)$ is an additive quantum error correcting code with parameters $[[n,k,d]]$.
\end{thm}

\subsection{Non-Additive Codes}
There are some codes whose structure may not be explained as the joint eigenspace of an Abelian subgroup of the error group. The first example of such a code was found in \cite{rains}, where the code was the direct sum of six one-dimensional stabilizer codes, or stabilizer states. These stabilizer states were related by translation using operators in the Pauli group. Further examples of such codes, using the Fourier transform for construction, were found in \cite{krp}. Nonadditive codes were also constructed in \cite{simple}, using an ad-hoc construction, and \cite{grassl}, using cosets of the normalizer to track translates of the stabilizer codes. Nonadditive codes were also constructed in \cite{graph1} using a graph-theoretical approach.

\subsection{Clifford Codes}
Clifford codes are a more general class of codes than stabilizer codes. They incorporate coding over higher dimensional systems (qudits), using error bases more general than the Pauli basis. This definition of Clifford codes is taken from \cite{clifford}.

\begin{dfn}
Let $E$ be an abstract error group with faithful irreducible ordinary representation $\rho$ of degree $\sqrt{E:Z(E)}$. Denote by $\phi$ the character of $S$ corresponding to this representation, that is $\phi(g) = Trace(\rho(g)) \forall g \in E$. Denote by $\chi$ an irreducible character of $N$ that is a constituent of the restriction of the character $\phi$ to $N$. Then the Clifford code with data $(E, \rho, N, \chi)$ is defined as the image of the orthogonal projector
\[
P=\frac{\chi(1)}{|N|} \sum_{n \in N} \chi(n^{-1}) \rho (n).
\]

To characterize the error correcting poperties of a Clifford code, the inertia subgroup $T(\chi)$ of the character $\chi$ is defined by
\[
T(\chi) = \left\lbrace g \in E| \chi(gxg^{-1}) = \chi(x) \forall x ]in N\right\rbrace.
\]
\end{dfn}
The qausi-kernel is defined as
\[
Z(\nu)= \left\lbrace n \in T||\nu(n)| =nu(1)  \right\rbrace.
\]
The following theorem results.
\begin{thm}\cite{clifford}
Let $Q$ be a Clifford code with data $\left( E, \rho, N, \chi \right)$. Denote by $\nu$ the irreducible character of $T(\chi)$ descibed above. The code $Q$ is able to correct a set of errors $S \subset E$ if and only if the conditions $s_{1}^{-1}s_{2} \notin T(\chi) \backslash Z(\nu)$ holds $\forall s_{1}, s_{2} \in S$.
\end{thm}

\chapter{The Fourier Transform Approach}

\begin{figure}
\includegraphics[scale=0.75]{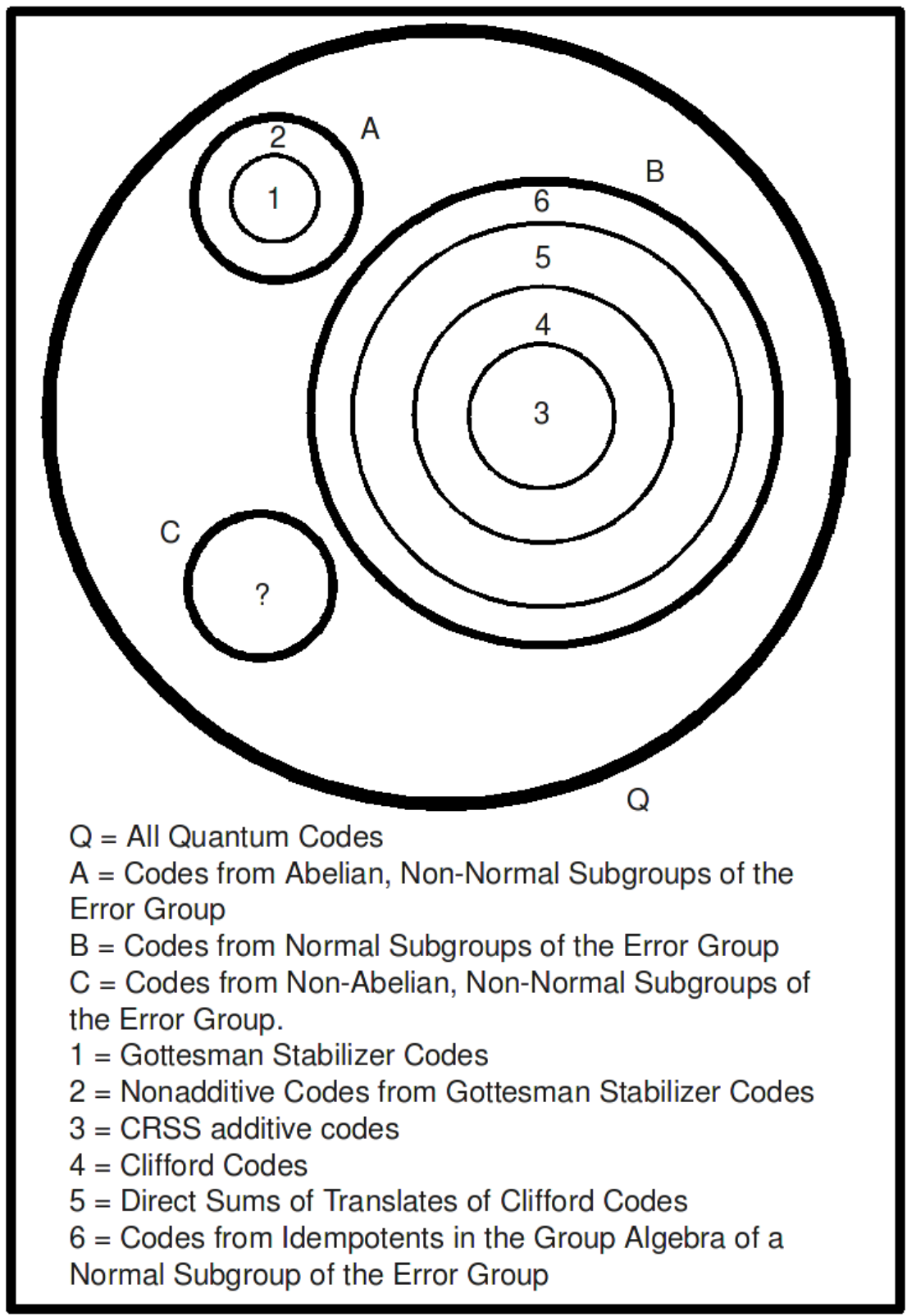} 
\caption{Venn Diagram Showing the Classes of Quantum Codes Obtainable by Fourier Inversion}
\end{figure}

\section{Introduction}
In this chapter, the Fourier transform over finite groups is introduced, and it is showed how it may be used to construct the various classes of quantum codes introduced in the previous chapter.

\section{The Fourier Inversion Formula and the Classes of Quantum Codes}

Consider a quantum system with $N$ levels. Let $G$ be a group of order $N^{2}$, with identity element $1$. We recall the definition of a nice error basis. A nice error basis \cite{beyond1} on $H=\mathbb{C}^{N}$ is a set $\left\lbrace \rho(g) \in U(N)|g\in G \right\rbrace $ such that
(i) $\rho(1)$ is the identity matrix
(ii) Trace$(\rho(g))$ = $n\delta_{g,1} \forall g \in G$
(iii) $\rho(g)\rho(h)=\omega(g,h) \rho(gh)$ $ \forall g,h \in G$ with $\omega(g,h) \in \mathbb{C}$. The $\omega$-covering of the nice error basis is called the abstract error group (or error group) of the nice error basis \cite{beyond1}. The error group modulo the center is called the index group of the nice error basis.

We fix a nice error basis on $n$ qudits (the associated Hilbert space being $H=\mathbb{C}^{d ^{\otimes n}}$). We form the associated error group, $E$. We choose $S$ a subgroup of $E$. In the group algebra, $\mathbb{C}S$, a typical element is $T=\sum_{s\in S} T_{s}s$. Say the irreducible representations of $S$ are given by $R=\left\lbrace \rho_{i}\right\rbrace $ and let Irr($S$)$=\left\lbrace \chi_{\rho_{i}}\right\rbrace $ denote the set of irreducible characters of $S$.

We recall the definition of the generalized Fourier transforms over a finite group $S$, and the Fourier inversion formula, following \cite{fft,serre}.

Given a finite group $S$, let $\rho_{i}:S \longrightarrow GL(W_{i}), \rho_{i}\in R$ be the distinct irreducible representations of $S$, upto isomorphism, and set $n_{i} = dim (W_{i})$. Each isomorphism

\begin{equation}
\Phi:\mathbb{C}S \longrightarrow \bigoplus_{i=1}^{i=k} \mathbb{C}^{n_{i}\times n_{i} }
\end{equation}

\noindent between the group algebra $\mathbb{C}S$ and the components $\mathbb{C}^{ n_{i} \times n_{i} }$  (known as the Wedderburn components \cite{fft}), is called a Fourier transform of the group $S$. A particular isomorphism is fixed by picking a system $\left\lbrace \rho_{1},...\rho_{k}\right\rbrace$ of representatives of irreducible representations of $S$, and defining $\Phi$ as the linear extension of the mapping $s \longrightarrow \bigoplus_{i=1}^{k} \rho_{i}(s), s \in S$.
 
The Fourier inversion formula \cite{serre} is given by

\begin{equation}
T_{s}=\frac{1}{|S|}\sum_{\rho_{i} \in R} n_{i}  Trace (\rho_{i}(s^{-1})a_{i}).
\end{equation}

\noindent where $n_{i} = Trace(\rho_{i}(1))$ is the dimension of the representation $\rho_{i}$, and $(a_{i})_{i \in \left\lbrace 1,\ldots k \right\rbrace }$ denotes the transform domain components. We note that each $a_{i}$ lies in $\mathbb{C}^{{n_{i}}\times n_{i}}$. We now study the inversion formula (2) and how it is related to quantum codes. Substituting the formula (2) into the expression $T=\sum_{s\in S} T_{s}s$, we have

\begin{equation}
T=\frac{1}{|S|}\sum_{s \in S} s\sum_{\rho_{i} \in R} n_{i} Trace (\rho_{i}(s^{-1})a_{i}).
\end{equation}

We constrain $a_{i} \in \mathbb{C}^{n_{i}\times n_{i}}$ such that $a_{i}^{2} = a_{i}$. Since convolution in the group algebra maps to pointwise multiplication in the transform domain, this allows us to find operators $T \in \mathbb{C} S$ with $T^{2}=T$. These $T$, being matrices, are projectors onto subspaces of the Hilbert space $H$, and hence represent quantum codes.

We now study the projectors, or quantum codes, obtainable from Fourier inversion on different subgroups $S$ of the error group $E$, with examples. The different classes are represented in Fig. 3.1.

\subsection{Class A: S an Abelian, Non-Normal Subgroup of the Error Group}
Following \cite{krp}, we assume the error group to have Abelian index group. We define a "Gottesman" subgroup of the error group to be an Abelian subgroup of $E$, not containing any non-trivial multiple of identity $\omega I, \omega \neq 1$. In this case, $S$ is not normal in $E$. Since all the irreducible representations of the Abelian group $S$ are 1-dimensional, and there are $|S|$ of them \cite{serre}, the transform domain components are specified by a vector $A = (a_{i})_{i\in S}$. If we choose $A = (\delta_{i,1})$, we obtain from (3.2) the projector

\[
T_{1} = \frac{1}{|S|} \sum_{s \in S} s .
\]

This is the projector for the stabilizer code with stabilizer group $S$. This class of codes is represented by the set "1" in Fig. 3.1. The error correcting properties are characterized by the centralizer $Z(S)$ of this group in $E$ \cite{gottesma}. The $[[5,1,3]]$ perfect quantum error correcting code \cite{perfect} is an example of such a code. If we choose several $a_{i}$ equal to 1, it is possible to obtain codes that are the direct sums of translates of Image$(T_{1})$ \cite{krp}. The $((5,6,2))$ non-additive code of \cite{rains} may be constructed like this. Such codes are represented by set "2" in Fig 3.1.

\subsection{Class B: S a Normal Subgroup of the Error Group}

For $S$ an Abelian, normal subgroup of the Pauli group for qubit systems, and a single $a_{i}$ taken as $I$, we obtain the codes of \cite{crss}. (Set"3" in Fig. 3.1). In this case, not all the transform components $a_{i}$ yield nontrivial codes. We require the condition that $\chi_{\rho_{i}}(\omega I) = \omega$, which makes the character $\chi_{\rho_{i}}$ yield the eigenvalue of the code space for each operator. In the language of Clifford theory, characters with $\chi_{\rho_{i}}(\omega I) \neq \omega$ are not irreducible constituents of the restriction of the representation of the error group $E$ to $S$.

For the case of $S$ a normal (not necesarily Abelian) subgroup of the error group (not necessarily the Pauli group), and a single $a_{i}=I$ such that $P_{\chi_{\rho_{i}}}$ is not $0$, we obtain a Clifford code (Set "4" in Fig. 3.1). This is possibly a "true" Clifford code only if the index group of the error group is non-Abelian. If the index group of the error basis is Abelian, all the Clifford codes obtained are stabilizer codes \cite{beyond2}.

We now consider the following case: $S$ a normal, not necessarily Abelian subgroup of $E$ and more than one $a_{i}=I$.
For this case, from (2),

\[ T = \sum_{\rho_{i} \in R} \sum_{s\in S} \frac{1}{|S|}n_{i}Trace(\rho_{i}(s^{-1})a_{i})
\]

 \[
=\sum_{\rho_{i} \in R, a_{i}=I}\sum_{s \in S}\frac{1}{|S|}n_{i}\chi_{\rho_{i}}(s^{-1})s
\]

\noindent where $\chi_{\rho_{i}}$ = Trace $\rho_{i}(s)$ is the irreducible character obtained from the irreducible representation $\rho_{i}$ of $S$.

We know from representation theory that 
\[
P_{\chi_{\rho_{i}}}=\sum_{s\in S} \frac{1}{|S|}n_{i}\chi_{\rho_{i}}(s^{-1})s
\]

\noindent is the projector onto the irreducible component of the space $\mathbb{C}^{d^{\otimes n}}$ associated with $\chi_{\rho_{i}} \in Irr(S)$. Hence,
\begin{equation}
T = \sum_{\rho_{i} \in R, a_{i}=1} P_{\chi_{\rho_{i}}}.
\end{equation}

We assume without loss of generality that each $P_{\chi_{\rho_{i}}}$ is non-zero (this is equivalent to the condition that $\rho_{i}$ is an irreducible constituent of the restriction of the representation of $E$ to $S$.) The projector $T$ is a sum of projectors for Clifford codes related by translation (this is a consequence of Clifford's theorem \cite{beyond2,charthry}, which says that the $\chi_{\rho_{i}}$ are all related by conjugation.) This has been explored for Pauli groups, but not for error groups with non-Abelian index groups. We compute the error detection properties of such codes (set "5" in Fig. 3.1) in the next section, and search for an example in the error groups available online at the Catalogue of Nice Error Bases (http://faculty.cs.tamu.edu/klappi/ueb/ueb.html).

We proceed now to the most general case in class B (set "6" in Fig. 3.1). In this case, we choose $a_{i}^{2}=a_{i}$. The $a_{i}$ need not be constrained to $\left\lbrace 0, I \right\rbrace$. For example,

\[
A = \left( 
\begin{array}{cc}
1&0 \\
0&0 \\
\end{array}
\right) 
\]

and 

\[
B = \left( 
\begin{array}{cc}
0.5&-0.5 \\
-0.5&0.5 \\
\end{array}
\right) 
\]

\noindent are valid values of $a_{i}$ for a two-dimensional representation $\rho_{i}$.

Although Clifford codes have been studied, and their error corecting properties characterized, this is not true for codes from "non-invertible idempotents in the transform domain" (like $A$ and $B$). In \cite{knill2}, mention is made of projectors obtained from "primitive orthogonal idempotents" of an irreducible character. These correspond to taking a single $a_{i}$ non-zero, and this $a_{i}$ having only a single diagonal element non-zero, and equal to 1 \cite{knill2}. This results in a code-space that is strictly smaller than a Clifford code. Setting several $a_{i}$ of this format, one can obtain the span of several such spaces as a code space. Most generally, one can use the $a_{i}$ satisfying $a_{i}^2=a_{i}$ and not necessarily invertible, to obtain code spaces. However, none of these has been systematically studied in the literature.

There are two cases to be considered: when the error group has Abelian index group, and when the error group has non-Abelian index group. An error group with Abelian index group forces set "4" to become set "3", and set "5" to become "direct sum of translates of additive codes" in Fig. 3.1. Using the most general idempotents in the transform domain, it may be possible to obtain new codes from error groups with Abelian index groups that are neither stabilizer codes nor the sums of their translates. We feel this justifies their further study.

\subsubsection{Error Group with Abelian Index Group}
We look for codes in set "6" of Fig. 3.1 when the index group is Abelian. As no general theory exists for such codes, we performed computer search on a small (5-qubit) Pauli group. We chose our group $S$ as follows: Set $G$ to be the stabilizer group descibing the $[[5,1,3]]$ perfect quantum error correcting code of \cite{perfect}. Choose $S$ to be the centralizer of $G$ in $E$. $S$ is normal in $E$, and has 80 irreducible representations, of which sixty-four are 1-dimensional, and sixteen 2-dimensional. Only the 2-dimensional representations contribute non-zero projectors, so we formed test sets loading transform values onto a subset of them. The values came from the set
\[
L =
\left\lbrace 0_{2}, I_{2}, P_{X+}, P_{X-}, P_{Y+}, P_{Y-}, P_{Z+}, P_{Z-}\right\rbrace ,
\]

\noindent where $P_{X+}$ denotes the 1-dimensional projector onto the $+1$ eigenspace of the Pauli $X$ operator, etc. We note that using $P_{Z+}$ or $P_{Z-}$ in a single location $a_{i}$ gives the "primitive orthogonal idempotents" of \cite{knill2}. Unlike the 1-dimensional case, there is an infinity of possible transform components (1 or 2-dimensional projectors) yielding projectors in the group algebra. It is an open problem as to how to design the transform domain projectors in order to get a good code space.

From our computer search, we observed that many, but not all, of the projectors in the group algebra $\mathbb{C}S$ are actually contained in $\mathbb{C}G$ where $G$ is an Abelian subgroup of $S$. We denote such codes by $A(S)$, i.e. codes that are obtainable from the group algebra of an Abelian subgroup of $S$. $A(S)$ contains only stabilizer codes or sums of their translates. All the Clifford codes of $S$ are in $A(S)$. Some non-Clifford codes of $S$ also turn up in $A(S)$. For our test group $S$, and test transform components, the results are tabulated in Table 1. We note that in this test-set, the codes outside $A(S)$ have nontrivial detectable sets, but perform very poorly in terms of minimum distance. The Clifford codes of $S$ are just stabilizer codes from $Z(S)$ (\cite{beyond2}, Theorem 6), and have minimum distance 3, as expected. 

\subsubsection{Error Group with Non-Abelian Index Group}
We begin by presenting some examples of the use of (2) for obtaining projectors (or quantum codes).

\subsection{Example Using Non-Invertible Non-Zero Transform Values}
An example of a non-invertible, non-zero transform component satisfying $a_{i}^{2}=a_{i}$ is given by
\[
a_{i}=A=\left( 
\begin{array}{cc}
1 & 0 \\
0 & 0 \\
\end{array}
\right) .
\]

Consider the error group on 4-level qudits, of size 32, with index group $C_{2}\times D_{8}$ \cite{clifford}. As demonstrated in \cite{clifford}, we can form the error group on one qudit and locate a normal subgroup within it that is isomorphic to a dihedral group with 16 elements. This dihedral group has four one dimensional representations and three two dimensional representations over $\mathbb{C}$ \cite{serre}. Taking the transform domain components to be $(0,0,0,0,A,0_{2},0_{2})$ where $A$ is taken as above, we obtain the projector

\[
T=\left( 
\begin{array}{cccc}
0&0&0&0 \\
0&0.5&0&-0.5i \\
0&0&0&0 \\
0&0.5i&0&0.5 \\

\end{array}
\right). 
\]

\noindent The one-dimensional image of $T$ can detect all errors in $E$.

\subsection{Example Using Only Invertible Non-Zero Transform Values}
Consider the same error group $E$ on 4-level qudits, with index group $C_{2}\times D_{8}$. We consider the same dihedral group of order 16 in $E$. Taking the transform domain components to be $(0,0,0,0,I_{2},0_{2},0_{2})$ we obtain, using the Fourier inversion formula, the projector

\[
T = 
\left( 
\begin{array}{cccc}
0.5&0&0.5i&0 \\
0&0.5&0&-0.5i \\
-0.5i&0&0.5&0 \\
0&0.5i&0&0.5 \\
\end{array}
\right). 
\]

This is the same projector obtained in \cite{clifford} using the tools of Clifford theory; it is the smallest example of a Clifford code that is not a stabilizer code. We have tabulated all the projectors available from non-Abelian normal subgroups of this error group in Tables 4.2-4.8. The transform values are taken from the same set $L=\left\lbrace 0_{2}, I_{2}, P_{X+}, P_{X-}, P_{Y+}, P_{Y-}, P_{Z+}, P_{Z-}\right\rbrace $. Hence, the tables include codes obtained using only the set $\left\lbrace 0, I\right\rbrace $, using only noninvertible non-zero transform values (like $P_{Z_{+}}$) and using a combination of invertible and non-invertible non-zero values in the transform domain. In making this table, we have considered only those transform components that contribute non-zero projectors (i.e. transform components corresponding to the irreducible constituents.) Similar tables can be generated for the other error groups available in the Catalogue of Nice Error Bases.

\section{Sums of Translates of Clifford Codes}
We assume henceforth that $a_{i}\in \left\lbrace 0, I \right\rbrace $. For this restricted case, we have seen (3.4) that

\[
T = \sum_{\rho_{i} \in R, a_{i}=1} P_{\chi_{\rho_{i}}},
\]

\noindent where 
\[
P_{\chi_{\rho_{i}}}=\sum_{s\in S} \frac{1}{|S|}n_{i}\chi_{\rho_{i}}(s^{-1})s
\]

\noindent is the projector onto the irreducible component of the space $\mathbb{C}^{d^{\otimes n}}$ associated with $\chi_{\rho_{i}} \in Irr(S)$. The code becomes a sum of translates of Clifford codes, for the $a_{i}$ chosen as above. We now treat the error detection properties of such codes. We assume without loss of generality that each $P_{\chi_{\rho_{i}}}$ is non-zero (this is equivalent to the condition that $\rho_{i}$ is an irreducible constituent of the restriction of the representation of $E$ to $S$.)

The projector $T$ is a sum of projectors for Clifford codes. By Clifford's theorem \cite{charthry}, the characters of the irreducible constitutents of the representation of S are related by conjugation.

We can rewrite (4) as $T=P_{\chi_{1}}+P_{\chi_{2}} + \ldots + P_{\chi_{t}}$. Then, Clifford's theorem says that
\[
  \chi_{i} = \chi_{1}^{h_{i}}
\]
\noindent for some $h_{i} \in E$. Here $\chi^{h}(s)=\chi (hsh^{-1})$ for some $h \in E$. We define the set $B = \left\lbrace h_{i} \right\rbrace$. We denote the image of the space $P_{\chi_{1}}$ by $W$, and the translates of $W$ by $hW, h \in B$. The $hW$ are the images of the $P_{\chi_{1}^{h}}$. We denote the quasikernel of $hW$ by $Z(hW)$, and the inertia subgroup of $W$ by $T(W)$ \cite{beyond2}. If $W$ is obtained as the image of the projector $P_{\chi}$, we can compute the inertia group from the character $\chi$. We denote this as $T(\chi)$.

\subsection{Error Detection Properties of Direct Sums of Translates of Clifford Codes}
We focus on error detection conditions for these codes. The error detection conditions for a quantum code \cite{knill2} state that an error $g$ can be detected by a code with projector $T$ if, and only if
\[
TgT = \phi (g)T
\]

\noindent for some $\phi (g) \in \mathbb{C}$. We consider three separate cases of $g\in E$.
\subsubsection*{Case 1: $g\in \cap Z(hW)$}
We recall the definition of $T(W)$ and $Z(W)$, the inertia subgroup and quasikernel \cite{beyond2} of the Clifford code $W$.

\[
T(W)=\left\lbrace g\in E |gW \tilde{=} W \right\rbrace.
\]

$Z(W)$ is defined to be the set of elements that act on the Clifford code $Q$ by scalar multiplication.
\[
Z(W) = \left\lbrace g\in T(W) | \exists \lambda\in \mathbb{C} \forall v\in Q, gv=\lambda v \right\rbrace.
\]

In this case, $g$ acts by scalar multiplication on each space $hW$. We have
\[
TgT = g.g^{-1}TgT
=g.(g^{-1}\sum_{h\in B} P_{\chi^{h}} g) T
\]
\[
=gTT = gT
=\sum_{h\in B} gP_{\chi^{h}}
\]
\[
=\sum_{h \in B} \lambda_{g}(h)P_{\chi^{h}} = \phi(g) \sum P_{\chi^{h}}.
\]
Equating the two sides, a necessary and sufficient condition for error detection is
\[
\forall h\in B, \lambda_{g}(h) = constant = \phi(g).
\]

It can be shown that this is a generalization of the first error correcting condition of \cite{krp} (though the subgroup $S$ taken there is a Gottesman subgroup, and not normal in the error group.)

\subsubsection*{Case 2: $g \notin S$}
In this case,
\[
TgT = g.g^{-1}TgT
\]
\[
=g.g^{-1}(\sum_{h\in B} P_{\chi^{h}} )g.T\]
\[
=g.\sum P_{\chi^{hg}}.T
\]
\[
=\phi(g).\sum_{h\in B} P_{\chi^{h}}
\]

We note that in the group algebra $\mathbb{C}E$, the LHS and RHS of the above equation have disjoint support, and hence, both sides must be zero for the error detection condition to hold.

\[
\sum_{h\in B} P_{\chi^{hg}}.T
=\sum_{h\in B} P_{\chi^{hg}} . \sum_{h' \in B} P_{\chi^{h'}}
\]

\[
=\sum_{h,h' \in B} P_{\chi^{hg}}.P_{\chi^{h^{'}}}
\]
\[
=\sum_{h,h'\in B} [\chi^{hg},\chi^{h^{'}}]P_{\chi^{h^{'}}}
\]
A necessary and sufficient condition for both sides of the equation to be zero is
\[
[\chi^{hg},\chi^{h^{'}}] = 0 \forall h,h^{'} \in B.
\]
\subsubsection*{Case 3:$g\in S, g \notin \cap Z(hW)$}
We note that $S$ is a subgroup of the inertia groups $T(\chi^{h}),  \forall h\in B$. We are therefore considering elements of the inertia groups that do not act by scalar multiplication on the code space. As showed in \cite{beyond2}, such errors cannot be detected by Clifford codes, hence cannot be detected by direct sums of Clifford codes either.

\section{Example of Sums of Translates of Clifford Codes}
It is hard to find non-trivial examples of direct sums of translates of Clifford codes, as all the Clifford codes known to us consist of coding on a single qudit \cite{clifford,remarks}. Hence the number of dimensions available for packing in translates is less. We consult \cite{remarks} for the following example. It may be noted that the only other possibility of a direct sum of Clifford codes in this table is the sum of two  2-dimensional Clifford codes in a 6-dimensional space, or the sum of two 3-dimensional Clifford codes in a 9-dimensional space. Both of these yield trivial detectable sets, however.

\subsection{Example: 2 Dimensional Projectors in 8 Dimensional Space}

We consider coding using $n=1, d=8$. We use the error basis $E$ with non-Abelian index group SmallGroup(64,10). A normal, non-Abelian subgroup $S$ is obtained in $E$ of size 32, and yields four projectors of dimension 2 each. We take the code $C$ to be the sum of the images of two of these projectors, $P_{1}$ and $P_{2}$.

\[P_{1} = \left( 
\begin{array}{cccccccc}
1&0&0&0&0&0&0&0\\
0&0&0&0&0&0&0&0\\
0&0&0&0&0&0&0&0\\
0&0&0&0&0&0&0&0\\
0&0&0&0&1&0&0&0\\
0&0&0&0&0&0&0&0\\
0&0&0&0&0&0&0&0\\
0&0&0&0&0&0&0&0\\
\end{array}
\right) 
\]

\[P_{2} = \left( 
\begin{array}{cccccccc}
0&0&0&0&0&0&0&0\\
0&1&0&0&0&0&0&0\\
0&0&0&0&0&0&0&0\\
0&0&0&0&0&0&0&0\\
0&0&0&0&0&0&0&0\\
0&0&0&0&0&1&0&0\\
0&0&0&0&0&0&0&0\\
0&0&0&0&0&0&0&0\\
\end{array}
\right) 
\]

For the code with projector $P=P_{1} + P_{2}$ (a direct sum of Clifford codes), we computed the detectable set using the computer algebra package GAP (Groups, Algorithms, Programming). This is a 68-element subset of the 128-element $E$. We verified manually that the error-detection conditions previously mentioned hold true. Examples of the three classes of errors from the error group are

\[
e_{1} = 
\left( 
\begin{array}{cccccccc}
-1 &  0 &  0 &  0 &  0 &  0 &  0 &  0 \\
 0 & -1 &  0 &  0 &  0 &  0 &  0 &  0 \\
 0 &  0 &  1 &  0 &  0 &  0 &  0 &  0 \\
 0 &  0 &  0 &  1 &  0 &  0 &  0 &  0 \\
 0 &  0 &  0 &  0 & -1 &  0 &  0 &  0 \\
 0 &  0 &  0 &  0 &  0 & -1 &  0 &  0 \\
 0 &  0 &  0 &  0 &  0 &  0 &  1 &  0 \\
 0 &  0 &  0 &  0 &  0 &  0 &  0 &  1 \\
\end{array}
\right) ,
\]

\[
e_{2}=
\left( 
\begin{array}{cccccccc}
0&0 &    -1  &    0 &     0  &    0  &    0   &   0 \\
0&0&0&-1&0&0&0&0 \\
0&  -i  &    0 &     0&      0 &     0 &     0  &    0 \\
-1&      0  &    0&      0&      0&      0&      0 &    0 \\
0  &    0    &  0&      0 &     0 &     0 &     0  &   -1 \\
0   &   0    &  0&      0 &     0 &     0 & -i  &    0 \\
0    &  0    &  0&      0 &  i &     0 &     0  &    0 \\
0    &  0    &  0&      0 &     0 &  i &     0  &    0 \\
\end{array}
\right) ,
\]

\[e_{3}=
\left( 
\begin{array}{cccccccc}
0&      0 &     0  &    0  &   -1 &     0 &     0 &     0 \\
0&      0 &     0  &    0  &    0 & -i &     0 &     0 \\
0&      0 &     0  &    0  &    0 &     0 &     0 &    -1 \\
0&      0 &     0  &    0  &    0 &     0 &    -1 &     0 \\
-1&      0&      0 &     0 &     0&      0&      0&      0 \\
0 &  i &     0  &    0  &    0 &     0 &     0 &     0 \\
0 &     0 &     0  &   -1  &    0 &     0 &     0 &     0 \\
0 &     0 &    -1  &    0  &    0 &     0 &     0 &     0 \\
\end{array}
\right) .
\]

We have chosen $e_{1}, e_{2}$ and $e_{3}$ corresponding to cases (1), (2) and (3) of the analysis for error detection. That is, $e_{1} \in \bigcap Z(hW)$, $e_{2} \notin S$ and $e_{3} \in S, e_{3} \notin \bigcap Z(hW)$. It is possible to verify that these errors are detectable, detectable and not detectable by the code with projector $P_{1}+P_{2}$, respectively.

\chapter{Computer Results}
\section{Introduction}
In this chapter, the computer search results are documented. Algorithms for Fourier transform and inversion, on subgroups of the Pauli group and more general abstract error groups, were implemented using the computer algebra package GAP (Groups, Algorithms, Programming).

\section{Computer Search Results}

In Table 4.1, the group $S$ was chosen as follows: Set $G$ to be the stabilizer group describing the $[[5,1,3]]$ perfect quantum error correcting code of \cite{perfect}. Choose $S$ to be the centralizer of $G$ in the Pauli group. The ``Transform components" column details the transform values that were loaded into a subset of the 2-dimensional representations of $S$. The ``In $A(S)$" column details whether the projector obtained has support exactly on an Abelian subgroup of S.``Is Clifford Code of $S$" column details whether the code obtained in a Clifford code of $S$. The remaining columns characterize the error detection/correction properties of the code obtained. The ``Size of detectable set" column details how many errors in the Pauli group can be detected. The remaining columns details how many weight 1 errors and weight 2 errors can be detected, and the minimum distance of the obtained code.

Tables 4.2 to 4.8 detail the search results for an error group with ID SmallGroup(32,16), having index group ID SmallGroup(16,3). These group IDs follow the convention of the GAP Catalogue of Small Groups. Various transform domain components are loaded and the detectable set is obtained for these.

\begin{table}[h]

\scriptsize
\begin{center}
\begin{tabular}[l]{|l|c|c|c|c|c|c|c|c|}
\hline
Sl.&Transform&Is Clifford &In $A(S)$?&Dimension&Size of &Wt. 1& Wt. 2&Minimum\\
&Components&Code of $S$?&&&detectable set&errors(of30)&errors(of180)&distance\\
\hline
1	&	$O_{2}$	,	$P_{Z-}$	&	No	&	Yes	&	1	&	2048	&	30	&	180	&	6	\\
2	&	$O_{2}$	,	$P_{X-}$	&	No	&	Yes	&	1	&	2048	&	30	&	180	&	6	\\
3	&	$O_{2}$	,	$P_{X+}$	&	No	&	Yes	&	1	&	2048	&	30	&	180	&	6	\\
4	&	$O_{2}$	,	$P_{Y+}$	&	No	&	Yes	&	1	&	2048	&	30	&	180	&	6	\\
5	&	$O_{2}$	,	$P_{Y-}$	&	No	&	Yes	&	1	&	2048	&	30	&	180	&	6	\\
6	&	$O_{2}$	,	$P_{Z+}$	&	No	&	Yes	&	1	&	2048	&	30	&	180	&	6	\\
7	&	$O_{2}$	,	$I_{2}$	&	Yes	&	Yes	&	2	&	1952	&	30	&	180	&	3	\\
8	&	$P_{Z-}$	,	$O_{2}$	&	No	&	Yes	&	1	&	2048	&	30	&	180	&	6	\\
9	&	$P_{Z-}$	,	$P_{Z-}$	&	No	&	Yes	&	2	&	1952	&	30	&	172	&	2	\\
10	&	$P_{Z-}$	,	$P_{X-}$	&	No	&	No	&	2	&	1840	&	28	&	168	&	1	\\
11	&	$P_{Z-}$	,	$P_{X+}$	&	No	&	No	&	2	&	1840	&	28	&	168	&	1	\\
12	&	$P_{Z-}$	,	$P_{Y+}$	&	No	&	No	&	2	&	1840	&	28	&	168	&	1	\\
13	&	$P_{Z-}$	,	$P_{Y-}$	&	No	&	No	&	2	&	1840	&	28	&	168	&	1	\\
14	&	$P_{Z-}$	,	$P_{Z+}$	&	No	&	Yes	&	3	&	1808	&	30	&	172	&	1	\\
15	&	$P_{Z-}$	,	$I_{2}$	&	No	&	Yes	&	3	&	1808	&	28	&	168	&	1	\\
16	&	$P_{X-}$	,	$O_{2}$	&	No	&	Yes	&	1	&	2048	&	30	&	180	&	6	\\
17	&	$P_{X-}$	,	$P_{Z-}$	&	No	&	No	&	2	&	1840	&	28	&	168	&	1	\\
18	&	$P_{X-}$	,	$P_{X-}$	&	No	&	Yes	&	2	&	1952	&	28	&	176	&	1	\\
19	&	$P_{X-}$	,	$P_{X+}$	&	No	&	Yes	&	2	&	1952	&	30	&	172	&	2	\\
20	&	$P_{X-}$	,	$P_{Y+}$	&	No	&	No	&	2	&	1840	&	28	&	168	&	1	\\
21	&	$P_{X-}$	,	$P_{Y-}$	&	No	&	No	&	2	&	1840	&	28	&	168	&	1	\\
22	&	$P_{X-}$	,	$P_{Z+}$	&	No	&	No	&	2	&	1840	&	28	&	168	&	1	\\
23	&	$P_{X-}$	,	$I_{2}$	&	No	&	Yes	&	3	&	1808	&	28	&	168	&	1	\\
24	&	$P_{X+}$	,	$O_{2}$	&	No	&	Yes	&	1	&	2048	&	30	&	180	&	6	\\
25	&	$P_{X+}$	,	$P_{Z-}$	&	No	&	No	&	2	&	1840	&	28	&	168	&	1	\\
26	&	$P_{X+}$	,	$P_{X-}$	&	No	&	Yes	&	2	&	1952	&	30	&	172	&	2	\\
27	&	$P_{X+}$	,	$P_{X+}$	&	No	&	Yes	&	2	&	1952	&	28	&	176	&	1	\\
28	&	$P_{X+}$	,	$P_{Y+}$	&	No	&	No	&	2	&	1840	&	28	&	168	&	1	\\
29	&	$P_{X+}$	,	$P_{Y-}$	&	No	&	No	&	2	&	1840	&	28	&	168	&	1	\\
30	&	$P_{X+}$	,	$P_{Z+}$	&	No	&	No	&	2	&	1840	&	28	&	168	&	1	\\
31	&	$P_{X+}$	,	$I_{2}$	&	No	&	Yes	&	3	&	1808	&	28	&	168	&	1	\\
32	&	$P_{Y+}$	,	$O_{2}$	&	No	&	Yes	&	1	&	2048	&	30	&	180	&	6	\\
33	&	$P_{Y+}$	,	$P_{Z-}$	&	No	&	No	&	2	&	1840	&	28	&	168	&	1	\\
34	&	$P_{Y+}$	,	$P_{X-}$	&	No	&	No	&	2	&	1840	&	28	&	168	&	1	\\
35	&	$P_{Y+}$	,	$P_{X+}$	&	No	&	No	&	2	&	1840	&	28	&	168	&	1	\\
36	&	$P_{Y+}$	,	$P_{Y+}$	&	No	&	Yes	&	2	&	1952	&	30	&	172	&	2	\\
37	&	$P_{Y+}$	,	$P_{Y-}$	&	No	&	Yes	&	2	&	1952	&	28	&	176	&	1	\\
38	&	$P_{Y+}$	,	$P_{Z+}$	&	No	&	No	&	2	&	1840	&	28	&	168	&	1	\\
39	&	$P_{Y+}$	,	$I_{2}$	&	No	&	Yes	&	3	&	1808	&	28	&	168	&	1	\\
40	&	$P_{Y-}$	,	$O_{2}$	&	No	&	Yes	&	1	&	2048	&	30	&	180	&	6	\\
41	&	$P_{Y-}$	,	$P_{Z-}$	&	No	&	No	&	2	&	1840	&	28	&	168	&	1	\\
42	&	$P_{Y-}$	,	$P_{X-}$	&	No	&	No	&	2	&	1840	&	28	&	168	&	1	\\
43	&	$P_{Y-}$	,	$P_{X+}$	&	No	&	No	&	2	&	1840	&	28	&	168	&	1	\\
44	&	$P_{Y-}$	,	$P_{Y+}$	&	No	&	Yes	&	2	&	1952	&	28	&	176	&	1	\\
45	&	$P_{Y-}$	,	$P_{Y-}$	&	No	&	Yes	&	2	&	1952	&	30	&	172	&	2	\\
46	&	$P_{Y-}$	,	$P_{Z+}$	&	No	&	No	&	2	&	1840	&	28	&	168	&	1	\\
47	&	$P_{Y-}$	,	$I_{2}$	&	No	&	Yes	&	3	&	1808	&	28	&	168	&	1	\\
48	&	$P_{Z+}$	,	$O_{2}$	&	No	&	Yes	&	1	&	2048	&	30	&	180	&	6	\\
49	&	$P_{Z+}$	,	$P_{Z-}$	&	No	&	Yes	&	2	&	1952	&	28	&	176	&	1	\\
50	&	$P_{Z+}$	,	$P_{X-}$	&	No	&	No	&	2	&	1840	&	28	&	168	&	1	\\
51	&	$P_{Z+}$	,	$P_{X+}$	&	No	&	No	&	2	&	1840	&	28	&	168	&	1	\\
52	&	$P_{Z+}$	,	$P_{Y+}$	&	No	&	No	&	2	&	1840	&	28	&	168	&	1	\\
53	&	$P_{Z+}$	,	$P_{Y-}$	&	No	&	No	&	2	&	1840	&	28	&	168	&	1	\\
54	&	$P_{Z+}$	,	$P_{Z+}$	&	No	&	Yes	&	2	&	1952	&	30	&	172	&	2	\\
55	&	$P_{Z+}$	,	$I_{2}$	&	No	&	Yes	&	3	&	1808	&	28	&	168	&	1	\\
56	&	$I_{2}$	,	$O_{2}$	&	Yes	&	Yes	&	2	&	1952	&	30	&	180	&	3	\\
57	&	$I_{2}$	,	$P_{Z-}$	&	No	&	Yes	&	3	&	1808	&	28	&	168	&	1	\\
58	&	$I_{2}$	,	$P_{X-}$	&	No	&	Yes	&	3	&	1808	&	28	&	168	&	1	\\
59	&	$I_{2}$	,	$P_{X+}$	&	No	&	Yes	&	3	&	1808	&	28	&	168	&	1	\\
60	&	$I_{2}$	,	$P_{Y+}$	&	No	&	Yes	&	3	&	1808	&	28	&	168	&	1	\\
61	&	$I_{2}$	,	$P_{Y-}$	&	No	&	Yes	&	3	&	1808	&	28	&	168	&	1	\\
62	&	$I_{2}$	,	$P_{Z+}$	&	No	&	Yes	&	3	&	1808	&	28	&	168	&	1	\\
63	&	$I_{2}$	,	$I_{2}$	&	Yes	&	Yes	&	4	&	1808	&	28	&	168	&	1	\\

\hline
\end{tabular}
\caption{Computer search results for codes from idempotents in the transform domain,for $S$}
\end{center}

\end{table}

\newpage

\begin{table}[h]
\begin{center}
Error group Id = [ 32, 6 ] \\
Index group Id = [ 16, 3 ] \\

\begin{tabular}{|c|c|c|c|c|}
\hline
Tx domain&Dimension&Size of \\
components&&detectable set\\
\hline
$I_{2}$&4&2\\
$P_{Z_{+}}$&2&10\\
$P_{Z_{-}}$&2&10\\
$P_{X_{+}}$&2&10\\
$P_{X_{-}}$&2&10\\
$P_{Y_{+}}$&2&20\\
$P_{Y_{-}}$&2&20\\
\hline
\end{tabular}
\caption{ Projectors from idempotents for subgroup with ID [8,3]}
\end{center}
\end{table}

\begin{table}[h]
\begin{center}
\begin{tabular}{|c|c|c|}
\hline
Tx domain&Dimension&Size of \\
components&&detectable set\\
\hline
 $I_{2}$&4&2\\
 $P_{Z_{+}}$&2&10\\
 $P_{Z_{-}}$&2&10\\
 $P_{X_{+}}$&2&10\\
 $P_{X_{-}}$&2&10\\
 $P_{Y_{+}}$&2&20\\
 $P_{Y_{-}}$&2&20\\
\hline
\end{tabular}
\caption{ Projectors from idempotents for subgroup with ID [8,4]}
\end{center}
\end{table}

\begin{table}
\scriptsize
\begin{center}
\begin{tabular}{|c|c|c|c|}
\hline
Tx domain components &Dimension&Size of detectable set \\

\hline
 $O_{2}$,$I_{2}$&2&12\\
 $O_{2}$,$P_{Z_{+}}$&1&32\\
 $O_{2}$,$P_{Z_{-}}$&1&32\\
 $O_{2}$,$P_{X_{+}}$&1&32\\
 $O_{2}$,$P_{X_{-}}$&1&32\\
 $O_{2}$,$P_{Y_{+}}$&1&32\\
 $O_{2}$,$P_{Y_{-}}$&1&32\\
 $I_{2}$,$O_{2}$&2&12\\
 $I_{2}$,$I_{2}$&4&2\\
 $I_{2}$,$P_{Z_{+}}$&3&2\\
 $I_{2}$,$P_{Z_{-}}$&3&2\\
 $I_{2}$,$P_{X_{+}}$&3&2\\
 $I_{2}$,$P_{X_{-}}$&3&2\\
 $I_{2}$,$P_{Y_{+}}$&3&2\\
 $I_{2}$,$P_{Y_{-}}$&3&2\\
 $P_{Z_{+}}$,$O_{2}$&1&32\\
 $P_{Z_{+}}$,$I_{2}$&3&2\\
 $P_{Z_{+}}$,$P_{Z_{+}}$&2&20\\
 $P_{Z_{+}}$,$P_{Z_{-}}$&2&12\\
 $P_{Z_{+}}$,$P_{X_{+}}$&2&2\\
 $P_{Z_{+}}$,$P_{X_{-}}$&2&2\\
 $P_{Z_{+}}$,$P_{Y_{+}}$&2&2\\
 $P_{Z_{+}}$,$P_{Y_{-}}$&2&2\\
 $P_{Z_{-}}$,$O_{2}$&1&32\\
 $P_{Z_{-}}$,$I_{2}$&3&2\\
 $P_{Z_{-}}$,$P_{Z_{+}}$&2&12\\
 $P_{Z_{-}}$,$P_{Z_{-}}$&2&20\\
 $P_{Z_{-}}$,$P_{X_{+}}$&2&2\\
 $P_{Z_{-}}$,$P_{X_{-}}$&2&2\\
 $P_{Z_{-}}$,$P_{Y_{+}}$&2&2\\
 $P_{Z_{-}}$,$P_{Y_{-}}$&2&2\\
 $P_{X_{+}}$,$O_{2}$&1&32\\
 $P_{X_{+}}$,$I_{2}$&3&2\\
 $P_{X_{+}}$,$P_{Z_{+}}$&2&2\\
 $P_{X_{+}}$,$P_{Z_{-}}$&2&2\\
 $P_{X_{+}}$,$P_{X_{+}}$&2&6\\
 $P_{X_{+}}$,$P_{X_{-}}$&2&6\\
 $P_{X_{+}}$,$P_{Y_{+}}$&2&10\\
 $P_{X_{+}}$,$P_{Y_{-}}$&2&10\\
 $P_{X_{-}}$,$O_{2}$&1&32\\
 $P_{X_{-}}$,$I_{2}$&3&2\\
 $P_{X_{-}}$,$P_{Z_{+}}$&2&2\\
 $P_{X_{-}}$,$P_{Z_{-}}$&2&2\\
 $P_{X_{-}}$,$P_{X_{+}}$&2&6\\
 $P_{X_{-}}$,$P_{X_{-}}$&2&6\\
 $P_{X_{-}}$,$P_{Y_{+}}$&2&10\\
 $P_{X_{-}}$,$P_{Y_{-}}$&2&10\\
 $P_{Y_{+}}$,$O_{2}$&1&32\\
 $P_{Y_{+}}$,$I_{2}$&3&2\\
 $P_{Y_{+}}$,$P_{Z_{+}}$&2&2\\
 $P_{Y_{+}}$,$P_{Z_{-}}$&2&2\\
 $P_{Y_{+}}$,$P_{X_{+}}$&2&10\\
 $P_{Y_{+}}$,$P_{X_{-}}$&2&10\\
 $P_{Y_{+}}$,$P_{Y_{+}}$&2&6\\
 $P_{Y_{+}}$,$P_{Y_{-}}$&2&6\\
 $P_{Y_{-}}$,$O_{2}$&1&32\\
 $P_{Y_{-}}$,$I_{2}$&3&2\\
 $P_{Y_{-}}$,$P_{Z_{+}}$&2&2\\
 $P_{Y_{-}}$,$P_{Z_{-}}$&2&2\\
 $P_{Y_{-}}$,$P_{X_{+}}$&2&10\\
 $P_{Y_{-}}$,$P_{X_{-}}$&2&10\\
 $P_{Y_{-}}$,$P_{Y_{+}}$&2&6\\
 $P_{Y_{-}}$,$P_{Y_{-}}$&2&6\\
\hline
\end{tabular}
\caption{ Projectors from idempotents for subgroup with ID [16,13]}
\end{center}
\end{table}

\begin{table}
\scriptsize
\begin{center}
\begin{tabular}{|c|c|c|c}
\hline
Tx domain components&Dimension&Size of detectable set\\

\hline
 $O_{2}$,$I_{2}$&2&12\\
 $O_{2}$,$P_{Z_{+}}$&1&32\\
 $O_{2}$,$P_{Z_{-}}$&1&32\\
 $O_{2}$,$P_{X_{+}}$&1&32\\
 $O_{2}$,$P_{X_{-}}$&1&32\\
 $O_{2}$,$P_{Y_{+}}$&1&32\\
 $O_{2}$,$P_{Y_{-}}$&1&32\\
 $I_{2}$,$O_{2}$&2&12\\
 $I_{2}$,$I_{2}$&4&2\\
 $I_{2}$,$P_{Z_{+}}$&3&2\\
 $I_{2}$,$P_{Z_{-}}$&3&2\\
 $I_{2}$,$P_{X_{+}}$&3&2\\
 $I_{2}$,$P_{X_{-}}$&3&2\\
 $I_{2}$,$P_{Y_{+}}$&3&2\\
 $I_{2}$,$P_{Y_{-}}$&3&2\\
 $P_{Z_{+}}$,$O_{2}$&1&32\\
 $P_{Z_{+}}$,$I_{2}$&3&2\\
 $P_{Z_{+}}$,$P_{Z_{+}}$&2&2\\
 $P_{Z_{+}}$,$P_{Z_{-}}$&2&2\\
 $P_{Z_{+}}$,$P_{X_{+}}$&2&2\\
 $P_{Z_{+}}$,$P_{X_{-}}$&2&2\\
 $P_{Z_{+}}$,$P_{Y_{+}}$&2&2\\
 $P_{Z_{+}}$,$P_{Y_{-}}$&2&2\\
 $P_{Z_{-}}$,$O_{2}$&1&32\\
 $P_{Z_{-}}$,$I_{2}$&3&2\\
 $P_{Z_{-}}$,$P_{Z_{+}}$&2&2\\
 $P_{Z_{-}}$,$P_{Z_{-}}$&2&2\\
 $P_{Z_{-}}$,$P_{X_{+}}$&2&2\\
 $P_{Z_{-}}$,$P_{X_{-}}$&2&2\\
 $P_{Z_{-}}$,$P_{Y_{+}}$&2&2\\
 $P_{Z_{-}}$,$P_{Y_{-}}$&2&2\\
 $P_{X_{+}}$,$O_{2}$&1&32\\
 $P_{X_{+}}$,$I_{2}$&3&2\\
 $P_{X_{+}}$,$P_{Z_{+}}$&2&2\\
 $P_{X_{+}}$,$P_{Z_{-}}$&2&2\\
 $P_{X_{+}}$,$P_{X_{+}}$&2&2\\
 $P_{X_{+}}$,$P_{X_{-}}$&2&2\\
 $P_{X_{+}}$,$P_{Y_{+}}$&2&2\\
 $P_{X_{+}}$,$P_{Y_{-}}$&2&2\\
 $P_{X_{-}}$,$O_{2}$&1&32\\
 $P_{X_{-}}$,$I_{2}$&3&2\\
 $P_{X_{-}}$,$P_{Z_{+}}$&2&2\\
 $P_{X_{-}}$,$P_{Z_{-}}$&2&2\\
 $P_{X_{-}}$,$P_{X_{+}}$&2&2\\
 $P_{X_{-}}$,$P_{X_{-}}$&2&2\\
 $P_{X_{-}}$,$P_{Y_{+}}$&2&2\\
 $P_{X_{-}}$,$P_{Y_{-}}$&2&2\\
 $P_{Y_{+}}$,$O_{2}$&1&32\\
 $P_{Y_{+}}$,$I_{2}$&3&2\\
 $P_{Y_{+}}$,$P_{Z_{+}}$&2&6\\
 $P_{Y_{+}}$,$P_{Z_{-}}$&2&6\\
 $P_{Y_{+}}$,$P_{X_{+}}$&2&6\\
 $P_{Y_{+}}$,$P_{X_{-}}$&2&6\\
 $P_{Y_{+}}$,$P_{Y_{+}}$&2&2\\
 $P_{Y_{+}}$,$P_{Y_{-}}$&2&2\\
 $P_{Y_{-}}$,$O_{2}$&1&32\\
 $P_{Y_{-}}$,$I_{2}$&3&2\\
 $P_{Y_{-}}$,$P_{Z_{+}}$&2&6\\
 $P_{Y_{-}}$,$P_{Z_{-}}$&2&6\\
 $P_{Y_{-}}$,$P_{X_{+}}$&2&6\\
 $P_{Y_{-}}$,$P_{X_{-}}$&2&6\\
 $P_{Y_{-}}$,$P_{Y_{+}}$&2&2\\
 $P_{Y_{-}}$,$P_{Y_{-}}$&2&2\\
 \hline
\end{tabular}
\caption{ Projectors from idempotents for subgroup with ID [16,8]}
\end{center}
\end{table}

\begin{table}
\scriptsize
\begin{center}
\begin{tabular}{|c|c|c|c|}
\hline
Tx domain components&Dimension&Size of detectable set\\

\hline
 $O_{2}$,$I_{2}$&2&12\\
 $O_{2}$,$P_{Z_{+}}$&1&32\\
 $O_{2}$,$P_{Z_{-}}$&1&32\\
 $O_{2}$,$P_{X_{+}}$&1&32\\
 $O_{2}$,$P_{X_{-}}$&1&32\\
 $O_{2}$,$P_{Y_{+}}$&1&32\\
 $O_{2}$,$P_{Y_{-}}$&1&32\\
 $I_{2}$,$O_{2}$&2&12\\
 $I_{2}$,$I_{2}$&4&2\\
 $I_{2}$,$P_{Z_{+}}$&3&2\\
 $I_{2}$,$P_{Z_{-}}$&3&2\\
 $I_{2}$,$P_{X_{+}}$&3&2\\
 $I_{2}$,$P_{X_{-}}$&3&2\\
 $I_{2}$,$P_{Y_{+}}$&3&2\\
 $I_{2}$,$P_{Y_{-}}$&3&2\\
 $P_{Z_{+}}$,$O_{2}$&1&32\\
 $P_{Z_{+}}$,$I_{2}$&3&2\\
 $P_{Z_{+}}$,$P_{Z_{+}}$&2&6\\
 $P_{Z_{+}}$,$P_{Z_{-}}$&2&6\\
 $P_{Z_{+}}$,$P_{X_{+}}$&2&10\\
 $P_{Z_{+}}$,$P_{X_{-}}$&2&10\\
 $P_{Z_{+}}$,$P_{Y_{+}}$&2&2\\
 $P_{Z_{+}}$,$P_{Y_{-}}$&2&2\\
 $P_{Z_{-}}$,$O_{2}$&1&32\\
 $P_{Z_{-}}$,$I_{2}$&3&2\\
 $P_{Z_{-}}$,$P_{Z_{+}}$&2&6\\
 $P_{Z_{-}}$,$P_{Z_{-}}$&2&6\\
 $P_{Z_{-}}$,$P_{X_{+}}$&2&10\\
 $P_{Z_{-}}$,$P_{X_{-}}$&2&10\\
 $P_{Z_{-}}$,$P_{Y_{+}}$&2&2\\
 $P_{Z_{-}}$,$P_{Y_{-}}$&2&2\\
 $P_{X_{+}}$,$O_{2}$&1&32\\
 $P_{X_{+}}$,$I_{2}$&3&2\\
 $P_{X_{+}}$,$P_{Z_{+}}$&2&10\\
 $P_{X_{+}}$,$P_{Z_{-}}$&2&10\\
 $P_{X_{+}}$,$P_{X_{+}}$&2&6\\
 $P_{X_{+}}$,$P_{X_{-}}$&2&6\\
 $P_{X_{+}}$,$P_{Y_{+}}$&2&2\\
 $P_{X_{+}}$,$P_{Y_{-}}$&2&2\\
 $P_{X_{-}}$,$O_{2}$&1&32\\
 $P_{X_{-}}$,$I_{2}$&3&2\\
 $P_{X_{-}}$,$P_{Z_{+}}$&2&10\\
 $P_{X_{-}}$,$P_{Z_{-}}$&2&10\\
 $P_{X_{-}}$,$P_{X_{+}}$&2&6\\
 $P_{X_{-}}$,$P_{X_{-}}$&2&6\\
 $P_{X_{-}}$,$P_{Y_{+}}$&2&2\\
 $P_{X_{-}}$,$P_{Y_{-}}$&2&2\\
 $P_{Y_{+}}$,$O_{2}$&1&32\\
 $P_{Y_{+}}$,$I_{2}$&3&2\\
 $P_{Y_{+}}$,$P_{Z_{+}}$&2&2\\
 $P_{Y_{+}}$,$P_{Z_{-}}$&2&2\\
 $P_{Y_{+}}$,$P_{X_{+}}$&2&2\\
 $P_{Y_{+}}$,$P_{X_{-}}$&2&2\\
 $P_{Y_{+}}$,$P_{Y_{+}}$&2&20\\
 $P_{Y_{+}}$,$P_{Y_{-}}$&2&12\\
 $P_{Y_{-}}$,$O_{2}$&1&32\\
 $P_{Y_{-}}$,$I_{2}$&3&2\\
 $P_{Y_{-}}$,$P_{Z_{+}}$&2&2\\
 $P_{Y_{-}}$,$P_{Z_{-}}$&2&2\\
 $P_{Y_{-}}$,$P_{X_{+}}$&2&2\\
 $P_{Y_{-}}$,$P_{X_{-}}$&2&2\\
 $P_{Y_{-}}$,$P_{Y_{+}}$&2&12\\
 $P_{Y_{-}}$,$P_{Y_{-}}$&2&20\\
 \hline
\end{tabular}
\caption{ Projectors from idempotents for subgroup with ID [16,7]}
\end{center}
\end{table}

\begin{table}
\scriptsize
\begin{center}
\begin{tabular}{|c|c|c|c|}
\hline
Tx domain components&Dimension&Size of detectable set\\

\hline
 $O_{2}$,$I_{2}$&2&20\\
 $O_{2}$,$P_{Z_{+}}$&1&32\\
 $O_{2}$,$P_{Z_{-}}$&1&32\\
 $O_{2}$,$P_{X_{+}}$&1&32\\
 $O_{2}$,$P_{X_{-}}$&1&32\\
 $O_{2}$,$P_{Y_{+}}$&1&32\\
 $O_{2}$,$P_{Y_{-}}$&1&32\\
 $I_{2}$,$O_{2}$&2&20\\
 $I_{2}$,$I_{2}$&4&2\\
 $I_{2}$,$P_{Z_{+}}$&3&2\\
 $I_{2}$,$P_{Z_{-}}$&3&2\\
 $I_{2}$,$P_{X_{+}}$&3&2\\
 $I_{2}$,$P_{X_{-}}$&3&2\\
 $I_{2}$,$P_{Y_{+}}$&3&2\\
 $I_{2}$,$P_{Y_{-}}$&3&2\\
 $P_{Z_{+}}$,$O_{2}$&1&32\\
 $P_{Z_{+}}$,$I_{2}$&3&2\\
 $P_{Z_{+}}$,$P_{Z_{+}}$&2&12\\
 $P_{Z_{+}}$,$P_{Z_{-}}$&2&12\\
 $P_{Z_{+}}$,$P_{X_{+}}$&2&2\\
 $P_{Z_{+}}$,$P_{X_{-}}$&2&2\\
 $P_{Z_{+}}$,$P_{Y_{+}}$&2&2\\
 $P_{Z_{+}}$,$P_{Y_{-}}$&2&2\\
 $P_{Z_{-}}$,$O_{2}$&1&32\\
 $P_{Z_{-}}$,$I_{2}$&3&2\\
 $P_{Z_{-}}$,$P_{Z_{+}}$&2&12\\
 $P_{Z_{-}}$,$P_{Z_{-}}$&2&12\\
 $P_{Z_{-}}$,$P_{X_{+}}$&2&2\\
 $P_{Z_{-}}$,$P_{X_{-}}$&2&2\\
 $P_{Z_{-}}$,$P_{Y_{+}}$&2&2\\
 $P_{Z_{-}}$,$P_{Y_{-}}$&2&2\\
 $P_{X_{+}}$,$O_{2}$&1&32\\
 $P_{X_{+}}$,$I_{2}$&3&2\\
 $P_{X_{+}}$,$P_{Z_{+}}$&2&2\\
 $P_{X_{+}}$,$P_{Z_{-}}$&2&2\\
 $P_{X_{+}}$,$P_{X_{+}}$&2&10\\
 $P_{X_{+}}$,$P_{X_{-}}$&2&10\\
 $P_{X_{+}}$,$P_{Y_{+}}$&2&10\\
 $P_{X_{+}}$,$P_{Y_{-}}$&2&10\\
 $P_{X_{-}}$,$O_{2}$&1&32\\
 $P_{X_{-}}$,$I_{2}$&3&2\\
 $P_{X_{-}}$,$P_{Z_{+}}$&2&2\\
 $P_{X_{-}}$,$P_{Z_{-}}$&2&2\\
 $P_{X_{-}}$,$P_{X_{+}}$&2&10\\
 $P_{X_{-}}$,$P_{X_{-}}$&2&10\\
 $P_{X_{-}}$,$P_{Y_{+}}$&2&10\\
 $P_{X_{-}}$,$P_{Y_{-}}$&2&10\\
 $P_{Y_{+}}$,$O_{2}$&1&32\\
 $P_{Y_{+}}$,$I_{2}$&3&2\\
 $P_{Y_{+}}$,$P_{Z_{+}}$&2&2\\
 $P_{Y_{+}}$,$P_{Z_{-}}$&2&2\\
 $P_{Y_{+}}$,$P_{X_{+}}$&2&10\\
 $P_{Y_{+}}$,$P_{X_{-}}$&2&10\\
 $P_{Y_{+}}$,$P_{Y_{+}}$&2&10\\
 $P_{Y_{+}}$,$P_{Y_{-}}$&2&10\\
 $P_{Y_{-}}$,$O_{2}$&1&32\\
 $P_{Y_{-}}$,$I_{2}$&3&2\\
 $P_{Y_{-}}$,$P_{Z_{+}}$&2&2\\
 $P_{Y_{-}}$,$P_{Z_{-}}$&2&2\\
 $P_{Y_{-}}$,$P_{X_{+}}$&2&10\\
 $P_{Y_{-}}$,$P_{X_{-}}$&2&10\\
 $P_{Y_{-}}$,$P_{Y_{+}}$&2&10\\
 $P_{Y_{-}}$,$P_{Y_{-}}$&2&10\\
 \hline
\end{tabular}
\caption{ Projectors from idempotents for subgroup with ID [16,6]}
\end{center}
\end{table}

\begin{table}
\scriptsize
\begin{center}
\begin{tabular}{|c|c|c|c|}
\hline
Tx domain components&Dimension&Size of detectable set\\

\hline
 $O_{2}$,$I_{2}$&2&12\\
 $O_{2}$,$P_{Z_{+}}$&1&32\\
 $O_{2}$,$P_{Z_{-}}$&1&32\\
 $O_{2}$,$P_{X_{+}}$&1&32\\
 $O_{2}$,$P_{X_{-}}$&1&32\\
 $O_{2}$,$P_{Y_{+}}$&1&32\\
 $O_{2}$,$P_{Y_{-}}$&1&32\\
 $I_{2}$,$O_{2}$&2&12\\
 $I_{2}$,$I_{2}$&4&2\\
 $I_{2}$,$P_{Z_{+}}$&3&2\\
 $I_{2}$,$P_{Z_{-}}$&3&2\\
 $I_{2}$,$P_{X_{+}}$&3&2\\
 $I_{2}$,$P_{X_{-}}$&3&2\\
 $I_{2}$,$P_{Y_{+}}$&3&2\\
 $I_{2}$,$P_{Y_{-}}$&3&2\\
 $P_{Z_{+}}$,$O_{2}$&1&32\\
 $P_{Z_{+}}$,$I_{2}$&3&2\\
 $P_{Z_{+}}$,$P_{Z_{+}}$&2&10\\
 $P_{Z_{+}}$,$P_{Z_{-}}$&2&10\\
 $P_{Z_{+}}$,$P_{X_{+}}$&2&6\\
 $P_{Z_{+}}$,$P_{X_{-}}$&2&6\\
 $P_{Z_{+}}$,$P_{Y_{+}}$&2&2\\
 $P_{Z_{+}}$,$P_{Y_{-}}$&2&2\\
 $P_{Z_{-}}$,$O_{2}$&1&32\\
 $P_{Z_{-}}$,$I_{2}$&3&2\\
 $P_{Z_{-}}$,$P_{Z_{+}}$&2&10\\
 $P_{Z_{-}}$,$P_{Z_{-}}$&2&10\\
 $P_{Z_{-}}$,$P_{X_{+}}$&2&6\\
 $P_{Z_{-}}$,$P_{X_{-}}$&2&6\\
 $P_{Z_{-}}$,$P_{Y_{+}}$&2&2\\
 $P_{Z_{-}}$,$P_{Y_{-}}$&2&2\\
 $P_{X_{+}}$,$O_{2}$&1&32\\
 $P_{X_{+}}$,$I_{2}$&3&2\\
 $P_{X_{+}}$,$P_{Z_{+}}$&2&6\\
 $P_{X_{+}}$,$P_{Z_{-}}$&2&6\\
 $P_{X_{+}}$,$P_{X_{+}}$&2&10\\
 $P_{X_{+}}$,$P_{X_{-}}$&2&10\\
 $P_{X_{+}}$,$P_{Y_{+}}$&2&2\\
 $P_{X_{+}}$,$P_{Y_{-}}$&2&2\\
 $P_{X_{-}}$,$O_{2}$&1&32\\
 $P_{X_{-}}$,$I_{2}$&3&2\\
 $P_{X_{-}}$,$P_{Z_{+}}$&2&6\\
 $P_{X_{-}}$,$P_{Z_{-}}$&2&6\\
 $P_{X_{-}}$,$P_{X_{+}}$&2&10\\
 $P_{X_{-}}$,$P_{X_{-}}$&2&10\\
 $P_{X_{-}}$,$P_{Y_{+}}$&2&2\\
 $P_{X_{-}}$,$P_{Y_{-}}$&2&2\\
 $P_{Y_{+}}$,$O_{2}$&1&32\\
 $P_{Y_{+}}$,$I_{2}$&3&2\\
 $P_{Y_{+}}$,$P_{Z_{+}}$&2&2\\
 $P_{Y_{+}}$,$P_{Z_{-}}$&2&2\\
 $P_{Y_{+}}$,$P_{X_{+}}$&2&2\\
 $P_{Y_{+}}$,$P_{X_{-}}$&2&2\\
 $P_{Y_{+}}$,$P_{Y_{+}}$&2&12\\
 $P_{Y_{+}}$,$P_{Y_{-}}$&2&20\\
 $P_{Y_{-}}$,$O_{2}$&1&32\\
 $P_{Y_{-}}$,$I_{2}$&3&2\\
 $P_{Y_{-}}$,$P_{Z_{+}}$&2&2\\
 $P_{Y_{-}}$,$P_{Z_{-}}$&2&2\\
 $P_{Y_{-}}$,$P_{X_{+}}$&2&2\\
 $P_{Y_{-}}$,$P_{X_{-}}$&2&2\\
 $P_{Y_{-}}$,$P_{Y_{+}}$&2&20\\
 $P_{Y_{-}}$,$P_{Y_{-}}$&2&12\\
 \hline
\end{tabular}
\caption{ Projectors from idempotents for subgroup with ID [16,11]}
\end{center}
\end{table}

\newpage
\onecolumn

\chapter{Concluding Remarks}

In this thesis, the basics of quantum mechanics as applied to quantum error correction have been presented. The various classes of active error control codes have been described, The Fourier transform has been introduced as a general tool to construct these classes of codes, and also another class of codes not yet developed in the literature (Class 6 in Fig. 3.1). 

\section{Scope for Future Work}
\begin{itemize}
\item A class of quantum error control codes outside the scope of this thesis is Operator Quantum Error Correction (OQEC). This is a eans of combined active error correction and passive error avoidance that includes the methods of Decoherence-Free Subspaces (DFS) and Noiseless Subsystems (NS) as special cases. As the name suggests, OQEC relies heavily on operator theory for its results. An interesting extension of this work would be generalize the Fourier Transform approach to OQEC and the methods of passive error control.
\item Another area for future work would be to search systematically for good codes within class 6 of Fig. 3.1, that is, codes that are obtained from idempotents in the transform domain
\item A possible area of future work would be to study the Knill-Laflamme conditions in the transform domain. A partial characterization, for the case of non-additive codes obtained from error groups with Abelian index groups, is obtained in \cite{krp} in terms of character theory. It would be interesting to generalize this to the other classes of quantum codes in Fig. 3.1.
\end{itemize}

\bibliographystyle{ref_thesis}
\bibliography{biblio}
  
\end{document}